%% file: main.tex
\documentclass[sigconf,authorversion,nonacm]{acmart}

\usepackage{booktabs} 

\usepackage{xcolor}
\input{macros}
\usepackage{url}
\usepackage{pgfplots}

\floatstyle{plain}
\restylefloat{figure}

\usepackage{tikz} 

\usepgfplotslibrary{groupplots}

\AtBeginDocument{%
  \providecommand\BibTeX{{%
    Bib\TeX}}}

\setcopyright{acmcopyright}
\copyrightyear{2023}
\acmYear{2023}
\acmDOI{XXXXXXX.XXXXXXX}

\acmPrice{15.00}
\acmISBN{978-1-4503-XXXX-X/18/06}

\begin{document}

\title{Simulating Quantum Computations on Classical Machines: A Survey}

\ccsdesc[500]{General and reference~Surveys and overviews}
\ccsdesc[500]{Computing methodologies~Modeling and simulation}
\ccsdesc[300]{Theory of computation~Quantum computation theory}
\ccsdesc[100]{Computing methodologies~Quantum mechanic simulation}
\ccsdesc[300]{Hardware~Quantum computation}
\ccsdesc[100]{Computer systems organization~Quantum computing}

\author{Kieran~Young}
\email{kdyoung@mtu.edu}
\affiliation{%
    \institution{Michigan Technological University}
    \streetaddress{1400 Townsend Drive}
    \city{Houghton}
    \state{Michigan}
    \country{USA}
    \postcode{49931}
}
\author{Marcus~Scese}
\email{mdscese@mtu.edu}
\affiliation{%
    \institution{Michigan Technological University}
    \streetaddress{1400 Townsend Drive}
    \city{Houghton}
    \state{Michigan}
    \country{USA}
    \postcode{49931}
}
\author{Ali~Ebnenasir}
\email{aebnenas@mtu.edu}
\affiliation{%
    \institution{Michigan Technological University}
    \streetaddress{1400 Townsend Drive}
    \city{Houghton}
    \state{Michigan}
    \country{USA}
    \postcode{49931}
}



\maketitle


\input{abstract}
\input{intro}

\input{prelim}
\input{schrodinger}
\input{decision}
\input{vqa}
\input{tnc}

\input{feynman}

\input{hybrid}

\input{heisenberg}

\input{framework}

\input{challenges}

\input{concl}
\ \\


\bibliography{biblio}

\end{document}

%% file: macros.tex
\usepackage{amsmath}

\usepackage{amssymb}

\usepackage{amsthm}
\usepackage{amsfonts}
\usepackage{graphicx}
\usepackage{mathtools}
\usepackage{hyperref}
\usepackage{xspace}
\usepackage{aliascnt}

\usepackage{tikz}
\usetikzlibrary{arrows}
\usetikzlibrary{shapes}
\usetikzlibrary{backgrounds}

\usepackage{epsfig}

\newcommand{\RecEnumLang}{\ensuremath{\Sigma_1^0}}
\newcommand{\CoRecEnumLang}{\ensuremath{\Pi_1^0}}

\newcommand{\SemiDec}[1][]{\RecEnumLang\ifx\empty#1\empty\else-#1\fi\xspace}
\newcommand{\CoSemiDec}[1][]{\CoRecEnumLang\ifx\empty#1\empty\else-#1\fi\xspace}

\mathchardef\breakingcomma\mathcode`\,
{\catcode`,=\active
 \gdef,{\breakingcomma\discretionary{}{}{}}
}

\hypersetup{
 pdfborderstyle={/S/U/W 0.5}
}

\usepackage{paralist}

\newenvironment{itemize*}%
{\begin{compactitem}}%
{\end{compactitem}}

\newenvironment{enumerate*}%
{\begin{compactenum}}%
{\end{compactenum}}

\usepackage{floatflt}

\usepackage{algorithm}
\usepackage{algorithmic}
\usepackage{subfig}

\newcommand{\mynewtheorem}[2]{
 \newaliascnt{#1}{theorem}
 \newtheorem{#1}[#1]{#2}
 \aliascntresetthe{#1}
 \expandafter\def\csname #1autorefname\endcsname{#2}
}

\mynewtheorem{lemma}{Lemma}
\mynewtheorem{corollary}{Corollary}
\mynewtheorem{example}{Example}
\theoremstyle{definition}
\mynewtheorem{problem}{Problem}
\mynewtheorem{definition}{Definition}
\mynewtheorem{remark}{Remark}
\mynewtheorem{observation}{Observation}

\usepackage{float}
\floatstyle{boxed}
\restylefloat{figure}

\newcommand{\ket}[1]{|#1\rangle}
\newcommand{\bra}[1]{\langle #1|}
\newcommand{\braket}[2]{\langle #1|#2\rangle}
\newcommand{\ketbra}[2]{\ket{#1}\bra{#2}}
\newcommand{\braketm}[3]{\langle #1|#2|#3\rangle}
\newcommand{\tr}[1]{\text{tr}(#1)}

%% file: abstract.tex

\noindent{\bf Abstract}. We present a comprehensive study of quantum simulation methods and quantum simulators for classical computers. We first study an exhaustive set of 150+ simulators and quantum libraries. Then, we short-list the simulators that are actively maintained and enable simulation of quantum algorithms for more than 10 qubits. As a result, we realize that most efficient and actively maintained simulators have been developed after 2010. We also provide a taxonomy of the most important simulation methods, namely Schrodinger-based, Feynman path integrals, Heisenberg-based, and hybrid methods. We observe that most simulators fall in the category of Schrodinger-based approaches. However, there are a few efficient simulators belonging to other categories. We also make note that quantum frameworks form their own class of software tools that provide more flexibility for algorithm designers with a choice of simulators/simulation method. Another contribution of this study includes the use and classification of optimization methods used in a variety of simulators. We observe that some state-of-the-art simulators utilize a combination of software and hardware optimization techniques to scale up the simulation of quantum circuits. We summarize this study by providing a roadmap for future research that can further enhance the use of quantum simulators in education and research.

%% file: intro.tex
\section{Introduction}
\label{sec:intro}

This paper presents a comprehensive survey on the state-of-the-art quantum simulation methods and simulators as important components of the quantum software stack. Quantum Simulators (QS) play a crucial role in advancing the field of Quantum Computing (QC).  In the absence of scalable quantum computers and the unavailability of quantum machines to the public,  quantum simulators are important tools towards designing and evaluating Quantum Algorithms (QAs), and comparing the noise level of the results of real-world quantum computers with simulation results.
As such, the evaluation of quantum algorithms on the quantum simulators that run on classical machines is the only option for broadening the participation of research and education communities in QC. Moreover, the use of QS helps in advancing knowledge in QC and in other disciplines (e.g., computational Physics and Chemistry). Thus, QS can help in (i) educating the work force that is highly needed by QC industry, (ii) analyzing quantum algorithms’ behavior when the number of qubits grows beyond the reach of current quantum machines, and (iii) providing a platform for experimental evaluation of quantum algorithms/circuits. As a result, understanding the landscape of quantum simulators has multiple benefits for researchers, engineers, and educators in terms of the challenges and opportunities of using and developing quantum simulators.   





Existing studies broadly focus on quantum computing and its applications, and provide cursory reviews on quantum simulators as one class of quantum software. For example, there are many surveys that study and classify quantum programming languages \cite{selinger2004brief,gay2006quantum,sofge2008survey,pakin2018survey,heim2020quantum,garhwal2021quantum}. 
Other works study the field of QC and its evolution in general \cite{upama2022evolution,gill2022quantum,hota2022taxonomy}. Some researchers have studied the development of Quantum Software Engineering (QSE) as a field \cite{zhao2020quantum,piattini2022quantum}. The roles of different types of tools such as compilers and simulators have been studied in other literature surveys \cite{sofge2008survey,chong2017programming,cruz2022quantum}. Several other surveys have classified the mutual contributions of QC and other fields such as security  \cite{bruss2007quantum,faruk2022review}, Machine Learning (ML) \cite{biamonte2017quantum,ramezani2020machine}, health care \cite{rasool2022quantum} and applications of QC in optimization problems \cite{ying2010quantum,ajagekar2019quantum}.   
Researchers who study QS mostly adopt two approaches: either select a few QS and study them in depth (i.e., vertical study) or consider a long list of QS without going into details of their design (i.e., horizontal study). 
An example of a vertical study includes Avila {\it et al.} \cite{qsimsAvila2020} where the authors compare seven state-of-the-art simulators in terms of their potential for exploiting the processing power of multi-core CPUs and GPUs, and consider other factors such as  programming model, support for vectorization, use of cache memory,  and the use of shared memory.  
Heng {\it et al.} \cite{heng2020exploiting} also study a few approaches for simulating quantum computations on a sinlge GPU or on multiple GPUs.  A horizontal study of QS is included in \cite{gill2022quantum} where the authors provide a comprehensive overview of the entire QC area.

\begin{figure}[h]
\centering
\includegraphics[  scale= 0.24]{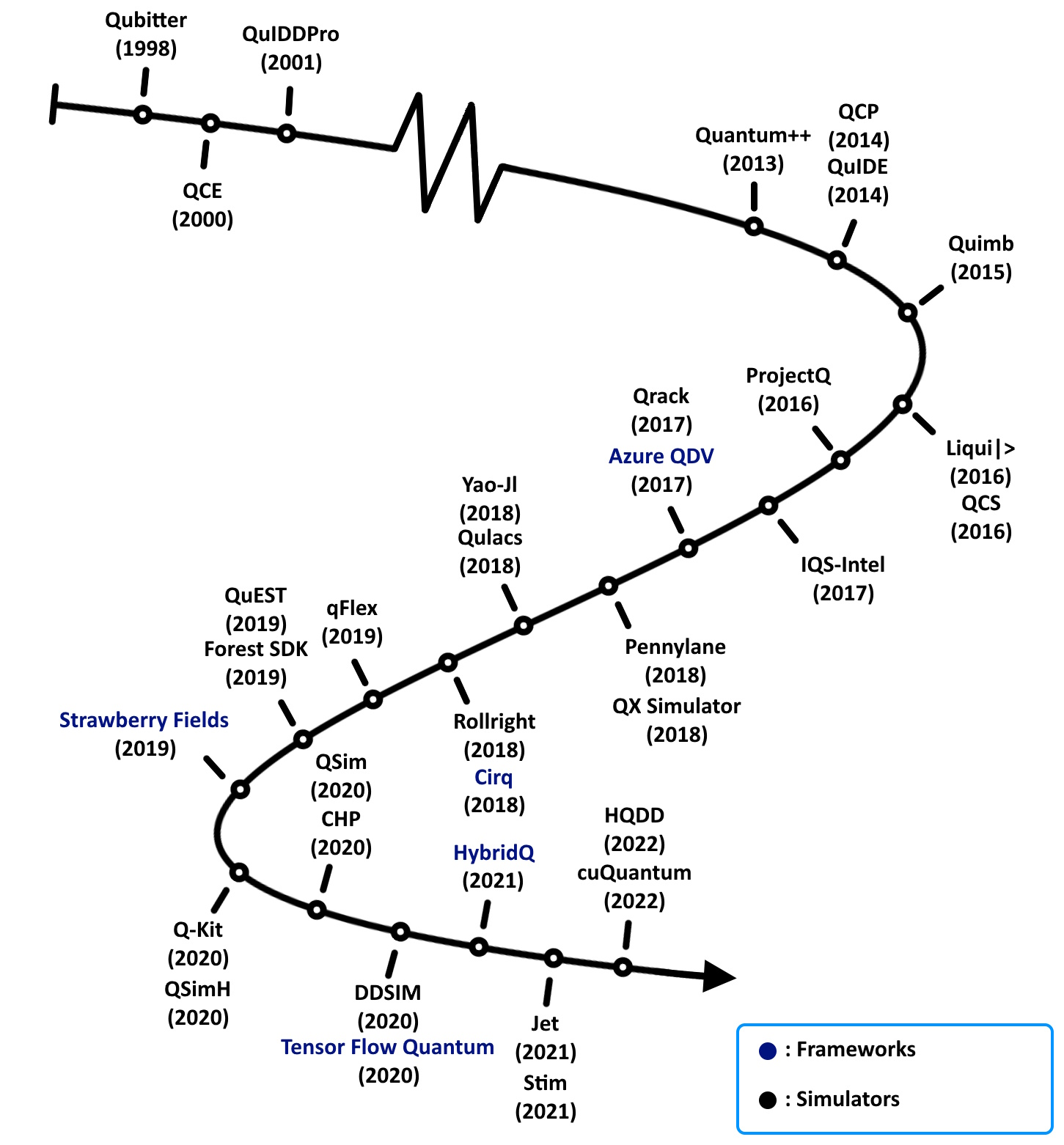}
\vspace*{-2mm}
\caption{\small Timeline of quantum computation simulators and frameworks.}
\label{fig:timeline}
\end{figure}

Contrary to most existing works, this paper provides a comprehensive survey on the landscape of QS based on important design criteria such as classic representation of quantum state, distribution of  quantum state on High Performance Computing (HPC) platforms, optimization methods for quantum operations, modern software and hardware technologies exploited in QS, and benchmarks and evaluation methods for QS.  Since simulators play a crucial role in both research and education, the main objective of this survey is to help researchers and educators to (i) learn more about the landscape of quantum simulators in the past decade, (ii) identify the right simulator for their application, and (iii) enhance their knowledge regarding the current challenges in the design and implementation of highly efficient quantum simulators.   To achieve this objective,  we conducted a two-step study of the literature. First, we started by an exhaustive search for existing quantum simulators, and pruned the initial list of more than 150 software based on the following criteria: Is it actively maintained and developed beyond 2010? Does it scale beyond 10 qubits on a regular office machine?
The output of the first step includes about 45 simulators that are actively maintained. These simulators have mostly been developed after 2010, illustrated in the timeline graph of Figure \ref{fig:timeline}. 
Quantum computer simulators have a rich and evolving history, with advancements in software tools and hardware leading to breakthroughs in the field. The earliest generalized quantum computer simulators were developed in the 1990s and as computers became more powerful, researchers began to develop more advanced quantum simulators that could handle larger, more complex QAs. 
With the emergence of quantum computers, researchers were presented with the immediate opportunity to simulate them. Figure 1 shows the rapid proliferation of quantum simulators with large qubit support, often achieved by limiting circuit depth or capability. In the early 2000s, much of the research focus shifted towards quantum algorithm and hardware development, resulting in a large gap in Figure 1. However, as hardware improved, the demand for more powerful simulators became evident and as progress slowed in one area, researchers turned their attention to classical solutions and began developing new quantum simulators capable of modeling various types of quantum computers. These new simulator types include decision diagrams, variational quantum algorithms, tensor network contraction, among others.

\begin{figure*}[h]
\centering
\includegraphics[  scale= 0.18]{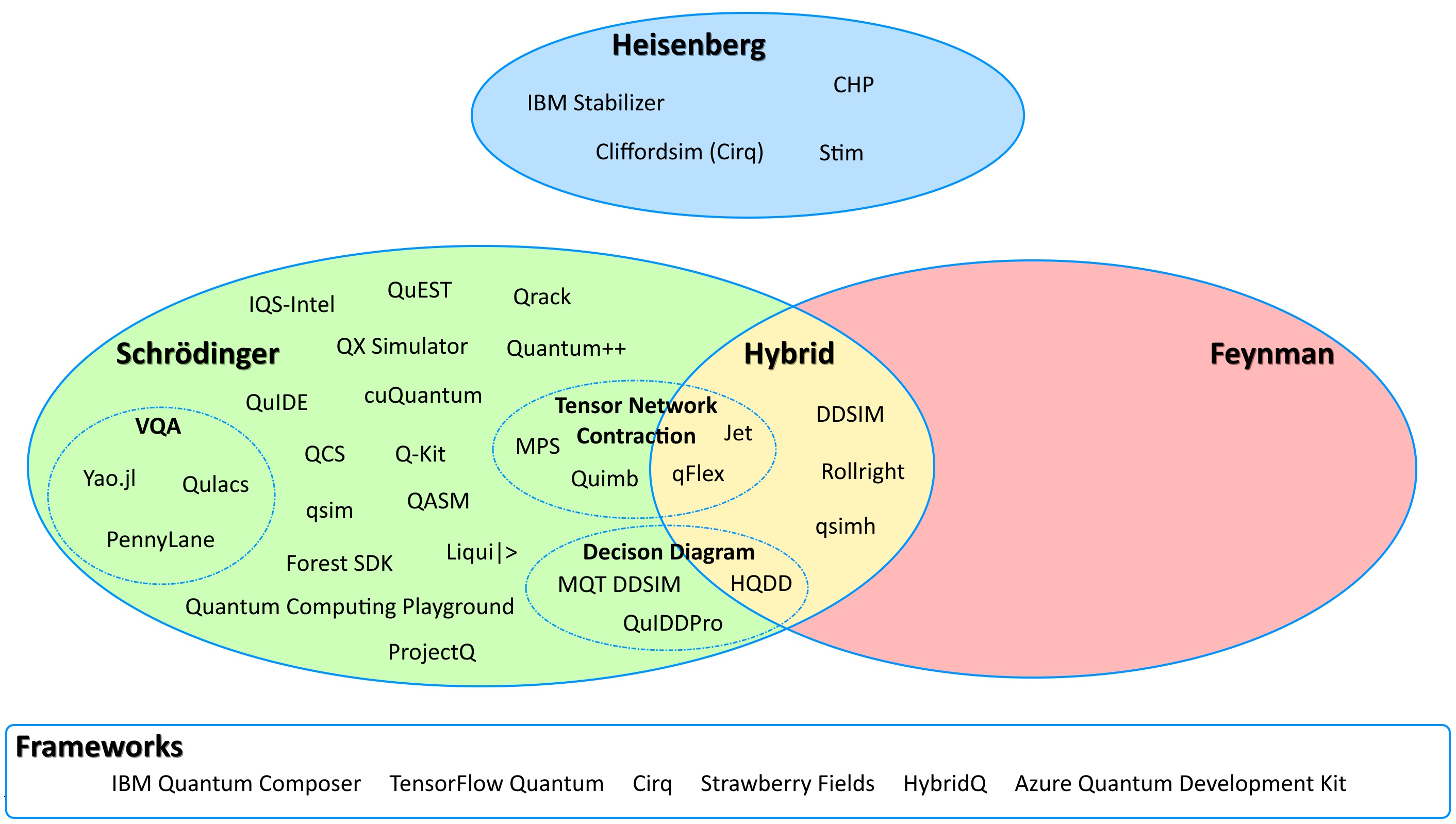}
\vspace*{-2mm}
\caption{\small Taxonomy of simulation methods and simulators.}
\label{fig:tax}
\end{figure*}

Second, we (i) classify these  state-of-the-art simulators for the research community based on their capabilities; (ii) identify the challenges of developing high-performance QS, and (iii) provide a roadmap for tackling the aforementioned challenges.  Figure \ref{fig:tax} illustrates the classification of  methods of simulating quantum computations on classical machines. We find that there exist four major approaches for the classical simulation of quantum computations, namely the Schr{\"o}dinger method (Section \ref{sec:schrod}), Feynman approach (Section \ref{sec:feynman}),  the Schr{\"o}dinger-Feynman hybrid approach (Section \ref{sec:hybrid}), and the Heisenberg method (Section \ref{sec:heisen}). We proceed by first providing some basic concepts of QC in Section \ref{sec:basics}. Then, in each one of the Sections \ref{sec:schrod} to \ref{sec:heisen} we study and classify the simulators under that approach based on how they model complex quantum states, optimize quantum state representation and transformations, utilize modern software/hardware resources, and evaluate the performance of simulators. Section \ref{sec:framework} discusses quantum frameworks (see the bottom of Figure \ref{fig:tax}) that provide high-level abstractions for the development, analysis, and simulation of quantum circuits.



%% file: prelim.tex
\section{Basics of Quantum Computing}
\label{sec:basics}

In this section, we represent a foundational understanding of quantum mechanics and quantum computing, drawn from various sources \cite{nielsen2000quantum, ebnenasir2022, quantikipuremixedstates, marinescu2012textbook, ekert2021kraus} which serves as the basis for this paper.
In the first Subsection (\ref{sec:fundamentals}), there is a review of fundamental concepts such as notation and qubits.
Following this, Subsection \ref{sec:qstate} expands on quantum states and their representations.
Finally, in Subsection \ref{sec:qops}, the requirements for quantum operations are detailed.

\subsection{Fundamentals of Quantum Mechanics}
\label{sec:fundamentals}
Quantum mechanics uses a particular notation that sometimes differs from the notation used in related fields: Dirac notation (Section \ref{sec:dirac}).
The most fundamental component necessary for quantum computation is the quantum bit or qubit as described in Section \ref{sec:qubits}.
These qubits interact through entanglement, a concept found in Section \ref{sec:entngl}.

\subsubsection{Dirac Notation}
\label{sec:dirac}
The Dirac notation \cite{marinescu2012textbook} provides a representation of basic quantum mechanics concepts in a complex vector space as follows:
\begin{itemize}
    \item The $\ket{\psi}$ is a column vector called a \textit{ket}.
    \item The $\bra{\psi}$ is a vector, called a \textit{bra}, which is the conjugate transpose to $\ket{\psi}$.
    \item The \textit{inner product} between two kets is represented as $\braket{\phi}{\psi}$ and as $\ketbra{\phi}{\psi}$ for the \textit{outer product}.
    The inner product between $\ket{\phi}$ and $A\ket{\psi}$ or $A^\dag \ket{\phi}$ and $\ket{\psi}$ is $\braketm{\phi}{A}{\psi}$ where $A$ is a square matrix.
    \item The $\oplus$ symbol represents the \textit{direct sum} of two vectors: $V \oplus W = (v_1, v_2) \oplus (w_1, w_2) = (v_1 + w_1, v_2 + w_2)$ \cite{reed1975physicsii}.
    \item The $\otimes$ symbol represents the \textit{tensor product}: a way of combining vector spaces to form a larger vector space.
    This is useful for multi-particle systems.
    Given the vectors $V$ and $W$ of dimension $m$ and $n$ respectively, $V \otimes W$ forms an $mn$ dimensional vector space.
    The Kronecker product matrix representation (Figure \ref{fig:tensorproductkronecker}) illustrates $A \otimes B$ where $A$ is an $m$ by $n$ matrix and $B$ is a $p$ by $q$ matrix.
    The resulting matrix is therefore $mp$ rows by $nq$ columns.
    \begin{figure}[h]
        \centering
        $A \otimes B \equiv
        \begin{bmatrix}
            A_{11}B & A_{12}B & \dots & A_{1n}B \\
            A_{21}B & A_{22}B & \dots & A_{2n}B \\
            \vdots & \vdots & \ddots & \vdots \\
            A_{m1}B & A_{m2}B & \dots & A_{mn}B
        \end{bmatrix}$
        \caption{Tensor Product Kronecker Product Representation}
        \label{fig:tensorproductkronecker}
    \end{figure}
    \item The \textit{complex conjugate transpose} (also known as the Hermitian transpose), is represented by the $\dag$ symbol on the matrix $A$ as $A^\dag = \begin{bmatrix} a & b \\ c & d \end{bmatrix}^\dag = \begin{bmatrix} a^* & c^* \\ b^* & d^* \end{bmatrix}$ where $a^*$ denotes the complex conjugate of the complex number $a$. 
    \item The \textit{trace} of a square matrix $A$ is the sum of its diagonal elements: $\tr{A} \equiv \sum_i A_{ii}$.
\end{itemize}

\subsubsection{Qubits}
\label{sec:qubits}
A qubit is an electron that represents a quantum bit of information \cite{ebnenasir2022}.
For example, an electron around a nucleus in the ground state could represent a 0 and a 1 in an excited state.
Due to the probabilistic nature of quantum mechanics, an electron is found to be in the ground or excited state with a certain probability after measurement.
Before measurement, the electron is in a \textit{superposition} of states: i.e., both states simultaneously.
Therefore, as expressed using Dirac notation, the electron is in the complex state $\alpha\ket{0} + \beta\ket{1}$ where $\alpha$ and $\beta$ are complex values \cite{nielsen2000quantum}.
Once measured, the superposition collapses and we can measure the qubit to be in state $\ket{0}$ with probability $|\alpha|^2$ or in $\ket{1}$ with probability $|\beta|^2$ where $|\alpha|^2 + |\beta|^2 = 1$.
In a 2-D vector space, $\alpha\ket{0} + \beta\ket{1}$ represents $\begin{pmatrix} \alpha\\\beta \end{pmatrix}$.
For example, $\alpha = 0$ and $\beta = 1$ results in $\alpha\ket{0} + \beta\ket{1} = \ket{1} = \begin{pmatrix} 0\\ 1 \end{pmatrix}$.
Note that after measurement, the system will always report the measured value since the superposition has collapsed into a single state.

The enumerated states of a classical computer with two bits are the states: $00$, $01$, $10$, and $11$. However, in quantum computation, two qubits have these four states in a combined, superposed state: $\ket{\phi} = \alpha_{00}\ket{00} + \alpha_{01}\ket{01} + \alpha_{10}\ket{10} + \alpha_{11}\ket{11}$. The normalized complex magnitude of the state would be $|\alpha_{00}|^2 + |\alpha_{01}|^2 + |\alpha_{10}|^2 + |\alpha_{11}|^2 = 1$ and the transposed vector would be $(\alpha_{00}, \alpha_{01}, \alpha_{10}, \alpha_{11})$.

Given two qubits $\ket{\phi} = \alpha_0\ket{0} + \alpha_1\ket{1}$ and $\ket{\psi} = \beta_0\ket{0} + \beta_1\ket{1}$, the two qubit system is the product of the two: $\alpha_0\beta_0\ket{00} + \alpha_0\beta_1\ket{01} + \alpha_1\beta_0\ket{10} + \alpha_1\beta_1\ket{11}$.
Measuring the $\ket{01}$ superposed state would simply result in the probability $|\alpha_0\beta_1|^2$ and the superposed state collapses into $\ket{01}$ with that probability.
Measuring $0$ in the first qubit has a probability of $|\alpha_0\beta_0|^2 + |\alpha_0\beta_1|^2$.
Since $0$ is measured in the first qubit, $10$ and $11$ become impossible states.
The superposed state of the system then becomes $\ket{\phi} = \alpha_0\beta_0 / \sqrt{|\alpha_0\beta_0|^2 + |\alpha_0\beta_1|^2}\ \ket{00} + \alpha_0\beta_1 / \sqrt{|\alpha_0\beta_0|^2 + |\alpha_0\beta_1|^2}\ \ket{01}$ such that the remaining probabilities sum to $1$ \cite{ebnenasir2022}.

\subsubsection{Entanglement}
\label{sec:entngl}
In the previous paragraph, after the first qubit was measured, the measurement of the second qubit seemed unrelated to that of the first \cite{ebnenasir2022}. This is not always the case due to quantum entanglement. If two quantum particles are close enough then they become entangled such that neither particle can be explained without the other. Consider the Bell state: $\ket{\phi} = \frac{1}{\sqrt{2}}\ket{00} + \frac{1}{\sqrt{2}}\ket{11}$. Note that the states $01$ and $10$ have probabilities of zero. Assume we measure the first qubit to be $0$. This invalidates the possibility of measuring the state $11$, so measuring the second qubit will undoubtedly result in $0$ with 100\% probability. When two quantum systems, such as two qubits, become entangled, measuring one qubit will collapse the quantum state and force a bit value on the other qubit. This phenomenon occurs regardless of distance \cite{ebnenasir2022}.

Furthermore, there are different levels of quantum entanglement.
A simple description has two levels: weak and strong entanglement.
A system of \textit{weakly} entangled particles can more easily be affected by its environment without the overall entanglement changing.
A system of \textit{strongly} entangled particles cannot share any entanglement with its environment.
Simply put, strongly entangled particles can withstand time and environment interactions better than weakly entangled particles.
In this way, a system of weakly entangled qubits have a lower computational potential \cite{cordelair2015entanglement}.

\subsection{Quantum States and Representations}
\label{sec:qstate}
Quantum states describe the properties of a quantum system.
As we are on the particle level, we consider the position, momentum, and spin of the particles.
These properties are described by the \textit{wave function}: the probability amplitudes of the quantum system encoded in complex values \cite{ebnenasir2022, marinescu2012textbook}.
One can visualize a qubit of that wave function by using the Bloch sphere (Section \ref{sec:bloch}).
Pure or mixed states (Section \ref{sec:pmstates}) are expressed depending on the knowledge of the system. 
For simulations, the wave function is usually expressed as a state vector or density matrix (Section \ref{sec:statedensity}) as a computational encoding of the wave function.
The wave functions are described as part of a vector space; often the Hilbert or Fock space (Section \ref{sec:hilbertfock}).

\subsubsection{Bloch Sphere}
\label{sec:bloch}
\begin{figure}[h]
    \centering
    \includegraphics[width=50mm]{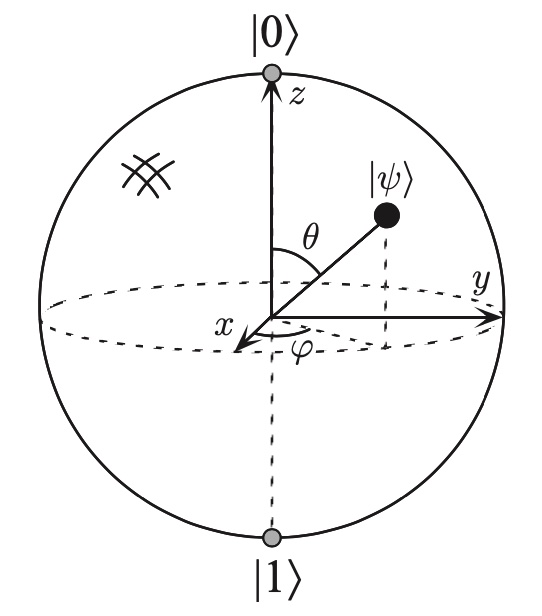}
    \caption{Bloch Sphere \cite{marinescu2012textbook}}
    \label{fig:blochsphere}
\end{figure}

The \textit{Bloch sphere} (Figure \ref{fig:blochsphere}) is a visualization technique for quantum state vectors \cite{marinescu2012textbook}.
Any point on or within this unit three-dimensional sphere describes a single qubit; however, there is no simple visualization with multiple qubits.
With $\ket{\psi} = \alpha \ket{0} + \beta \ket{1}$, a point $(\theta, \phi)$ is described by $\alpha = \cos(\theta / 2)$ and $\beta = e^{i \phi} \sin(\theta / 2)$ \cite{marinescu2012textbook}.

\subsubsection{Pure and Mixed States}
\label{sec:pmstates}
\textit{Pure states} are expressed as a single state vector with perfect knowledge of the system.
\textit{Mixed states} are characterized by a limitation in the knowledge of the state of a system; therefore, mixed states cannot be defined by a single state vector.
They are, however, defined as a mixture of pure states \cite{marinescu2012textbook}.
This uncertainty often increases computational complexity for simulating mixed states due to the requirement of multiple pure-state state vectors.
As a visualization technique, pure states are points on the surface of a Bloch sphere while mixed states are points on the interior of the sphere \cite{quantikipuremixedstates}.

\subsubsection{State Vectors and Density Matrices}
\label{sec:statedensity}
\textit{State vectors} are complex vectors that represent pure states in a Hilbert space.
The \textit{density matrix} (also known as the density operator) is an alternative formulation to the state vector which provides convenience in describing mixed states \cite{marinescu2012textbook}.
The density matrix can be written as $\rho \equiv \sum p_i \ket{\psi_i} \bra{\psi_i}$ where $\ket{\psi}$ denotes a pure state, and $p_i$ is an associate probability of being in the state $\ket{\psi}$.
In this way, the density matrix is an ensemble of pure state \cite{marinescu2012textbook}.
Therefore, pure states are commonly expressed as state vectors; and mixed states, as density matrices.
The density operator expression is simply $\rho = \ketbra{\psi}{\psi}$ for pure states \cite{marinescu2012textbook}.

\subsubsection{Hilbert Space and Fock Space}
\label{sec:hilbertfock}
Also known as an inner product space, a \textit{Hilbert space} is a vector space supporting the inner product of two vectors to output a complex number \cite{nielsen2000quantum}.
In Dirac notation, the inner product of two vectors $\psi$ and $\phi$ is expressed as $\braket{\psi}{\phi}$.
A Hilbert space enables superpositions through vectors.
Adding qubits to a Hilbert space increases the space exponentially, making this a challenge for simulations.
As the Hilbert space describes the state of a single particle using a vector, the Fock space describes a variable number of particles \cite{band2013textbook}.
That is akin to the sum of a set of Hilbert spaces:
$F_S(H) = \bigoplus^{\infty}_{n = 0} S_n H^{\otimes n} = \mathbb{C} \oplus H \oplus (S_n (H \otimes H)) \oplus (S_n (H \otimes H \otimes H)) \otimes \hdots$ where $S_n$ are the symmetrization tensor operators \cite{reed1975physicsii}.
The symmetrization operators create a symmetric or antisymmetric tensor.
For example, for $2\times 2$ matrices, the symmetrization operator is $T(A)=\frac{1}{2}(A + A^T)$ and the antisymmetrization operator is $T(A)=\frac{1}{2}(A - A^T)$ \cite{peyam2019symmetrization, reed1980physicsi}.

\subsection{Quantum Operations}
\label{sec:qops}
Quantum operations must be unitary (Section \ref{sec:unitary}).
However, there are non-unitary operations such as the measuring of a quantum system as described in Section \ref{sec:measurement}.
To facilitate such operations, the Kraus representation in Section \ref{sec:kraus} offers a modeling method for a quantum system.
Hamiltonians (Section \ref{sec:hamilton}) offer an alternative model by describing a quantum system in terms of energy dynamics over time.
To ultimately craft quantum circuits, quantum gates (Section \ref{sec:gates}) are applied to qubits, thereby manipulating the state vector and achieving computation.

\subsubsection{Unitary Operations}
\label{sec:unitary}
In the context of quantum computing, the operations that are applied on Hilbert spaces are often unitary.
Specifically, a matrix (or operator) $U$ is \textit{unitary} if and only if the following is satisfied: $U^{\dag}U = UU^{\dag} = I$.
This property allows for the reversibility of computation.
More crucially, a unitary operator preserves the probabilities of measurement over time.
Mathematically, a unitary operator $U$ preserves the inner products between vectors $\ket{v}$ and $\ket{w}$: $(U\ket{v}, U\ket{w}) = \braketm{v}{U^\dag U}{w} = \braketm{v}{I}{w} = \braket{v}{w}$.
A \textit{unitary transformation} preserves the norm of vectors while a \textit{unitary evolution} is described by a unitary transformation.
\cite{marinescu2012textbook}.

\subsubsection{Measurement}
\label{sec:measurement}
The action of measurement on a quantum system can be represented mathematically using projectors.
For example, with $\ket{\psi}$ the \textit{projector} is written as $P = \ketbra{\psi}{\psi}$, the outer product.
Projectors have several noteworthy properties that mirror quantum phenomenon.
Measuring a quantum system results in its collapse into the projector associated with that outcome.
This projector is orthogonal to the other possible outcomes, such that once measured, all other outcomes have a probability of zero.
Projectors also have the idempotent property where $P^2 = P$.
This property means that repeating the same measurement, or applying the projector twice, gives the same result as the first measurement or application of the projector \cite{marinescu2012textbook}.

Note that measurement is not affected by phase.
The \textit{phase}, more specifically in this context the \textit{global phase factor}, is the component of a quantum state that has no affect on observation.
For example, the state $e^{i\theta}\ket{\psi}$ is equal to $\ket{\psi}$ up to the phase $e^{i\theta}$ where $\ket{\psi}$ is a state vector and $\theta$ is a real number \cite{nielsen2000quantum}.

\subsubsection{The Kraus Representation}
\label{sec:kraus}
Quantum systems are often expressed using the Kraus representation which is also known as the operator-sum representation \cite{marinescu2012textbook}.
This representation is used when looking at quantum systems that are contained within an environment.
There are therefore two components of a quantum system: the \textit{principle system} and an \textit{environment}.
The principle system is the system of interest and the environment is a system that interacts with the principle system.
This interaction describes an \textit{open} quantum system while the two systems together form a \textit{closed} quantum system.
To access the principle system in isolation of the environment, we perform a \textit{partial trace}: an operation that leaves us with a reduced density matrix of the principle system while preserving the effects of entanglement with the environment \cite{marinescu2012textbook}.

Building up to the Kraus representation, start by looking at the principle system's state vector within the environment as they transform, or evolve, together over time: $\ket{\psi} \otimes \ket{e} \longrightarrow \sum_k E_k \ket{\psi} \otimes \ket{e_k}$ where $E_k$ is a Kraus operator and $\longrightarrow$ signifies the evolution \cite{ekert2021kraus}.
Kraus operators can be any set of linear operators that satisfy the completeness relation: $\sum_k E^\dag_k E_k = I$ \cite{marinescu2012textbook, ekert2021kraus}.
However, these Kraus operators are chosen to reflect the measurement process.
The completeness relation keeps the application of the Kraus operators unitary \cite{ekert2021kraus}. 

To see the impact of the entire evolution on the principle system, we start with the initial density operator (rewriting in terms of projectors): $\ketbra{\psi}{\psi} \otimes \ketbra{e}{e} \longrightarrow \sum_{k,l} E_k \ketbra{\psi}{\psi} E^\dag_l \otimes \ketbra{e_k}{e_l}$.
Then, by tracing over the environment, we isolate the principle system: $\ketbra{\psi}{\psi} \longrightarrow \sum_{k,l} E_k \ketbra{\psi}{\psi} E^\dag_l \braket{e_l}{e_k}$ where $\braket{e_l}{e_k}$ can be simplified to the Kronecker delta $\delta_{lk}$, finally resulting in $\sum_k E_k \ketbra{\psi}{\psi} E^\dag_k$.
Lastly, expressing in terms of the density operator leaves a compact expression called the \textit{Kraus representation} of the evolution of the open system: $\rho \longrightarrow \sum_k E_k \rho E^\dag_k$.
Without referring to the environment, the evolution of the principle system can therefore be concisely expressed as $\sum_k E_k \rho E^\dag_k$ where $\rho$ is the initial density operator of the system \cite{ekert2021kraus}.

\subsubsection{Hamiltonians}
\label{sec:hamilton}
In quantum mechanics, the Hamiltonian of a system serves as a mathematical operator that represents the total kinetic and potential energy of all particles within that system.
It not only characterizes the energy state of the system at a given moment, but also governs how this state evolves over time.
When it comes to problem-solving, the Hamiltonian is employed to model a problem in such a way that its ground state corresponds to a solution of that problem.
Unlike quantum circuits, which rely on sequences of gates acting on qubits, Hamiltonians are defined by parameters influenced by both the time and spatial properties of the system.
This unique formulation allows the energy levels and external influences encapsulated in the Hamiltonian to delineate the temporal evolution of the system, ultimately leading to valuable insights and potential solutions to complex problems \cite{ebnenasir2022, AmazonBraketAHS}.
The vast majority of QC simulators focus on gate-based computing rather than Hamiltonian-based, but the Hamiltonian method is indeed an alternative to consider.

\subsubsection{Quantum Gates}
\label{sec:gates}
Any unitary matrix is a valid quantum gate \cite{marinescu2012textbook}.
Single qubit gates are $2 \times 2$ matrices, and $4 \times 4$ matrices represent two qubit gates: with the vector $\ket{\psi} = \alpha \ket{0} + \beta \ket{1}$, gates must preserve $|\alpha|^2 + |\beta|^2 = 1$.
Figure \ref{fig:qgates} visualizes some common gates (described below) and their matrix representations.
\begin{itemize}
\item \textbf{I}: The identity matrix. Applying this gate does not alter the qubit.
\item \textbf{X}: Known as the "NOT" gate, the X gate transforms $\ket{\psi} = \alpha \ket{0} + \beta \ket{1}$ into $\beta \ket{0} + \alpha \ket{1}$.
This is a rotation in the Bloch sphere about the $\hat{x}$ axis of $\pi$ radians ($180^\circ$) \cite{marinescu2012textbook}.
\item \textbf{Y}: Starting with $\ket{\psi} = \alpha\ket{0} + \beta\ket{1}$, the Y gate creates $i\beta\ket{0} - i\alpha\ket{1}$.
This is a rotation in the Bloch sphere about the $\hat{y}$ axis of $\pi$ radians ($180^\circ$) \cite{marinescu2012textbook}.
\item \textbf{Z}: The Z gate flips $\ket{1}$ to $-\ket{1}$ and leaves $\ket{0}$ unchanged.
This is a rotation in the Bloch sphere about the $\hat{z}$ axis of $\pi$ radians ($180^\circ$) \cite{marinescu2012textbook}.
\item \textbf{H}: The Hadamard gate turns $\ket{0}$ into $(\ket{0} + \ket{1})/\sqrt{2}$ and $\ket{1}$ into $(\ket{0} - \ket{1})/\sqrt{2}$.
Also described as the "square-root of NOT" gate, the H gate turns $\ket{0}$ and $\ket{1}$ each halfway between $\ket{0}$ and $\ket{1}$.
Visually, on a Bloch sphere, the H gate is a rotation in the sphere with a $\pi/2$ radian ($90^\circ$) rotation about the $\hat{y}$ axis and $\pi$ radian ($180^\circ$) about the $\hat{x}$ axis.
Note that $H^2 = I$ since this is akin to a $2\pi$ radian ($360^\circ$) rotation \cite{marinescu2012textbook}.
\item \textbf{T}: After applying the T gate to $\ket{\psi} = \alpha\ket{0} + \beta\ket{1}$, the qubit becomes $\alpha\ket{0} + e^{i \pi / 4}\beta\ket{1}$.
This is a $\pi / 4$ radian ($45^\circ$) rotation around the $\hat{z}$ axis of the Bloch sphere \cite{marinescu2012textbook}.
\item \textbf{S}: The S gate converts $\ket{\psi} = \alpha\ket{0} + \beta\ket{1}$ to $\alpha\ket{0} + i\beta\ket{1}$. This is a rotation around the Bloch sphere about the $\hat{z}$ axis of $\pi / 2$ radians ($90^\circ)$ \cite{marinescu2012textbook}.
\item \textbf{SWAP}: The SWAP gate swaps two qubits such that the qubits $\ket{\psi} = \alpha \ket{0} + \beta \ket{1}$ and $\ket{\phi} = \gamma \ket{0} + \delta \ket{1}$ becomes $\ket{\psi} = \gamma \ket{0} + \delta \ket{1}$ and $\ket{\phi} = \alpha \ket{0} + \beta \ket{1}$ \cite{marinescu2012textbook}.
\item \textbf{Controlled NOT (CX)}: The CX gate takes two qubits, the control and the target. When the control qubit is $\ket{1}$ then the target qubit is flipped. Otherwise, neither qubit is altered \cite{marinescu2012textbook}.
\item \textbf{Controlled Phase Shift (CZ)}: The CZ gate also takes two qubits. 
If the control qubit and the target qubit are both $\ket{1}$, then the target qubit is transformed into $-\ket{1}$. Otherwise, there is no change.
\item \textbf{Controlled U (CU)}: The $U$ gate is an arbitrary quantum gate. Like the CX and CZ gates, other gates have \textit{controlled} variants. A control qubit controls whether or not $U$ is applied to the target qubit. If the control qubit is $\ket{1}$, then the target qubit is transformed from state $\ket{\psi}$ to $U\ket{\psi}$. Otherwise, the neither qubit is altered.
\item \textbf{$n$-qubit Controlled U (C$^n$U)}: Controlled gates can also have multiple control qubits and multiple source qubits where all qubits must be $\ket{1}$ for the target qubits to be affected.
\end{itemize}

Measurement can also be used in tandem with quantum gates.
Some gates are controlled by the measurement of one or more qubits rather than the qubits themselves.
This is known as a \textit{classical control} where the measured result of a qubit ($0$ or $1$) is used to decide whether or not to apply the gate to the target qubit \cite{zhou2000gateconstruction}.

\begin{figure*}[h]
\centering
$I \equiv \begin{bmatrix} 1 & 0 \\ 0 & 1 \end{bmatrix}$ \hspace{4mm}
$X \equiv \begin{bmatrix} 0 & 1 \\ 1 & 0 \end{bmatrix}$ \hspace{4mm}
$Y \equiv \begin{bmatrix} 0 & -i \\ i & 0 \end{bmatrix}$ \hspace{4mm}
$Z \equiv \begin{bmatrix} 1 & 0 \\ 0 & -1 \end{bmatrix}$ \hspace{4mm}
$H \equiv \frac{1}{\sqrt{2}} \begin{bmatrix} 1 & 1 \\ 1 & -1 \end{bmatrix}$ \hspace{4mm}
$T \equiv \begin{bmatrix} 1 & 0 \\ 0 & e^{i \pi / 4}\end{bmatrix}$ \hspace{4mm}
$S \equiv \begin{bmatrix} 1 & 0 \\ 0 & i \end{bmatrix}$ \hspace{4mm}
$SWAP \equiv \begin{bmatrix} 1 & 0 & 0 & 0 \\ 0 & 0 & 1 & 0 \\ 0 & 1 & 0 & 0 \\ 0 & 0 & 0 & 1 \end{bmatrix}$ \hspace{4mm}
$CX \equiv \begin{bmatrix} 1 & 0 & 0 & 0 \\ 0 & 1 & 0 & 0 \\ 0 & 0 & 0 & 1 \\ 0 & 0 & 1 & 0 \end{bmatrix}$ \hspace{4mm}
$CZ \equiv \begin{bmatrix} 1 & 0 & 0 & 0 \\ 0 & 1 & 0 & 0 \\ 0 & 0 & 1 & 0 \\ 0 & 0 & 0 & -1 \end{bmatrix}$ \hspace{4mm}
$CU \equiv \begin{bmatrix} I & 0 \\ 0 & U \end{bmatrix}$ \hspace{4mm}
\caption{Common Quantum Gates in Matrix Form}
\label{fig:qgates}
\end{figure*}

\subsection{Simulation Optimizations}
We summarize the known optimization techniques each simulator uses in Figures \ref{fig:schOpt}, \ref{fig:hybOpt}, and \ref{fig:heiOpt}.
As the information available with some simulators is limited, this list is incomplete; however, it provides a good overview of the separable optimizations we consider noteworthy.
\begin{itemize}
    \item \textbf{PAR}: The parallelization (PAR) optimizations include standard multi-threading (e.g. OpenMP), and vector or matrix partitioning.
    \item \textbf{SIMD}: The Single-Instruction, Multiple-Data (SIMD) optimization makes use of CPU hardware vector accelerators.
    Hardware implementations may have long registers suited to perform vector operations efficiently in parallel.
    Common SIMD implementations include AVX, FMA, and SSE \cite{faz2013simdsurvey}.
    \item \textbf{DECOMP}: In matrix decomposition (DECOMP), a matrix is broken down into a product of matricies, potentially enhancing computational efficiency or enabling parallel matrix operations.
    Matrix decomposition also includes the Schmidt decomposition, stabilizer decomposition, and low-rank stabilizer decomposition.
    \item \textbf{TRDLVL}: Thread-level parallelism (TRDLVL) is distinct from PAR-type parallelism in that thread-level is likely a lower-level, highly-intentional focus on parallelism.
    \item \textbf{LIB}: A linear-algebraic library (LIB) allows for simulators to utilize preexisting mathematical functions that were built for general purposes.
    \item \textbf{REDUC}: Precision reduction (REDUC) is an optimization focused on reducing the number of operations required to apply one or more gates.
    Consider the Hadamard gate: there are no imaginary components, and so the complex operations associated with the imaginary component can simply be skipped \cite{khammassi2017qx}.
    This optimization can be fruitfully combined with gate fusion (FUS).
    \item \textbf{FUS}: Gate fusion (FUS) and coalescing involves combining gates that are close spatially and temporally.
    Such fused gates are larger than their constituent gates; nevertheless, fewer overall operations are required, leading to speedup \cite{isakov2021simulations}.
    \item \textbf{TENS}: A tensor flow library (TENS) such as TensorFlow or TensorFlow Quantum enables efficient tensor operations especially with support for GPU acceleration.
    \item \textbf{KRAUS}: The Kraus representation (KRAUS) is a way of considering systems within an environment, as discussed in Section \ref{sec:kraus}.
    This representation is used by simulators to mathematically model quantum systems.
    \item \textbf{RECYC}: Qubit recycling (RECYC) involves reusing a qubit that has already been measured.
    Doing so reduces the required qubits, reducing the state vector size, reducing execution time \cite{tankasalaQuantum-Kit2019}.
\end{itemize}

%% file: schrodinger.tex
\section{Schr{\"o}dinger Method}
\label{sec:schrod}

The Schr{\"o}dinger method of simulation captures quantum states and unitary transformations as vectors and matrices respectively. In principle, this method of modeling is straightforward; however, such vectors and matrices contain an exponential number of $O(2^n)$ complex values for a circuit with $n$ input qubits. In practice, this becomes prohibitively expensive to represent in the memory of a classic machine. While there are techniques \cite{viamontes2007efficient,viamontes2009quantum,zulehner2018advanced} for mitigating this space complexity, the exponential cost is incurred in general. This section presents a thorough study of Schr{\"o}dinger-based simulation methods and simulators developed based on this approach. Specifically, Section \ref{sec:pureSchr} discusses methods that directly represent quantum vectors and matrices in a linear-algebraic way; i.e., {\it pure Schr{\"o}dinger}. Then, Section \ref{sec:decD} reviews methods that utilize decision diagrams for compact representation of vectors and matrices. Then, Section \ref{sec:vqa}  provides an overview of Variational Quantum Algorithms (VQA) methods for better utilization of our current Noisy Intermediate Scale Quantum (NISQ) machines. Finally, Section \ref{sec:tnc} provides an overview of some of the more basic tensor network contraction simulators on top of reviewing how this form of algorithm would work to calculate quantum states.


\subsection{Pure Schr{\"o}dinger}
\label{sec:pureSchr}
This section studies and classifies some of the state-of-the-art simulators that model quantum states and transformations in a linear-algebraic fashion in the memory of classic machines. 

\noindent{\bf IQS-Intel} \cite{guerreschi2020intel}: The Intel Quantum Simulator (IQS) utilizes High Performance Computing (HPC) and cloud platforms to enable a high performance simulator. IQS uses groups of resources of CPUs and GPUs in order to simulate distributed quantum states and to perform quantum computing operations on them. The simulation of quantum states is based on partitioning vectors of complex amplitudes and allocating each part to a different process. Complex quantum operations on large registers are decomposed to operations in single-qubit and controlled two-qubit circuits. Experimental evaluations of IQS indicate its performance on solving Quantum Approximate Optimization Algorithm (QAOA) for Max-Cut on 3-regular graphs, PSO, and QFT problems. They can simulate systems of up to 40 qubits for the aforementioned problems. IQS has been used as the backend of several programming environments (e.g., IBM's Qiskit, Google's Cirq, Microsoft's Azure, ProjectQ, Amazon's Braket).

\noindent{\bf QuEST} \cite{jones2019quest}: The Quantum Exact Simulation Toolkit (QuEST) is a standalone application that provides a highly efficient platform-independent simulator. The main design principle of QuEST is based on representing pure states and superpositions by vectors, and mixed quantum states by density matrices.  QuEST represents quantum circuits in terms of the universal gates $H, T, CZ, X^{1/2}$, and $Y^{1/2}$, namely the Hadamard,  $\pi/8$, controlled-phase and root Pauli $X$ and $Y$ gates.  QuEST is optimized for use in distributed and multithreaded platforms. QuEST supports both OpenMP and MPI-based parallelization and can be used in a hybrid way. A pure quantum state is captured by $2^n$ complex values, both the real and imaginary part in double precision. QuEST also utilizes instruction set vectorization and SIMD technology in order to speed up operations. The developers of QuEST experimentally evaluate it for pseudo random quantum circuits up to 38 qubits. QuEST provides a C-based language that runs on sequential, multithreaded, and parallel platforms transparent for the programmer. It also has a Mathematica interface called QuEST link, which supports circuit drawing.


\noindent{\bf Qrack} \cite{Qrack2017, strano2023qrack}:  Qrack is a simulator implemented in C++ and OpenCL for the simulation of quantum systems with at least 32 qubits. Qrack decomposes matrix operations to operations on $2\times 2$ matrices so such operations become highly parallelizable. The basic operations are performed using single-qubit, two-qubit, and three-qubit controlled gates. This design principle makes Qrack suitable for execution on HPC platforms. The developers of Qrack have experimentally evaluated it for QFT, random circuits rectangular Sycamore, single qubit and CCNOT (doubly-controlled NOT gate). Qrack can be used as a backend simulator in ProjectQ.


\noindent{\bf QX Simulator} \cite{khammassi2017qx}:  QX is a high-performance universal quantum computing simulator which simulates the execution of quantum circuits on perfect or error-prone quantum computers. QX simulates quantum computations on a classical supercomputer, called Lisa. Developers can use a QASM-based programming language, called Quantum Code, to write quantum algorithms on QX. Moreover, QX has a python interface. Quantum states are modeled in a linear-algebraic way. QX benefits from several optimizations including utilizing vector instructions of modern processors, reduction of floating point operations in sparse matrices, swap-based implementation where the result of multiplying permutation matrices/operators (e.g., Pauli-X) by a vector (i.e., quantum state) is obtained by swapping some elements of the vector, and exploiting thread-level parallelism. 
Experiments show that QX can simulate fully-entangled quantum systems up to 34 qubits. QX achieves 14 to 95 times speedup over Liqui$\mid>$ for QFT, entanglement, and Grover search algorithms. QX appears to be  a robust simulator, but it lacks sufficient  documentation.

\begin{figure*}[h]
\centering
\begin{tabular}{ |p{18mm}||p{7mm}|p{9mm}|p{16mm}|p{14mm}|p{7mm}|p{13mm}|p{6mm}|p{9mm}|p{13mm}|p{13mm}| }
 \hline
 \multicolumn{11}{|c|}{{\bf Optimization Techniques in Schr{\"o}dinger-based Simulators}} \\
 \hline
 {\bf Simulator} & {\bf PAR} & {\bf SIMD} & {\bf DECOMP}  &  {\bf TRDLVL}  &  {\bf LIB}  &  {\bf REDUC}  &  {\bf FUS}  &  {\bf TENS} &  {\bf KRAUS} & {\bf RECYC}\\
 \hline
 Intel-IQS   &  \checkmark    &  &  & & & & & &  & \\
  \hline
QuEST  &   \checkmark    & \checkmark  &  & & & & & & & \\
  \hline
Qrack  &   \checkmark      &  & \checkmark    & & & & & & & \\
  \hline
QX    &      &  \checkmark &   &  \checkmark& &  \checkmark & & &  & \\
  \hline
  Quantum++     &   \checkmark    &  &  & & \checkmark  & & & & & \\
  \hline
   LIQUi$\mid >$   &      &  &  & & & &  \checkmark  & &  & \\
  \hline
   cuQuantum   &      &  &  & & & &   & \checkmark &  & \\
  \hline
    Q-Kit     &      &  &  & & & &   & & &  \checkmark \\
  \hline
     DM1   &      &  &  & & & &   & & \checkmark & \\
  \hline
     Cirq   &      &  &  & & & &   & & \checkmark   & \\
  \hline
       Qualcs   &   \checkmark     & \checkmark   &  & & & &   & &  & \\
  \hline
qsim & \checkmark & \checkmark & & & & & \checkmark & \checkmark & & \\ \hline
Fock backend & \checkmark & & & & & & & & & \\ \hline
TensorFlow backend & & & & & & & & \checkmark & & \\ \hline
State Vector & \checkmark & \checkmark & & & & & & & & \\ \hline
cirq.Simulator & & & & & & & & & \checkmark & \\ \hline

ProjectQ & & & \checkmark & & & & & & & \\ \hline

PennyLane & & & & & & & & \checkmark & & \\ \hline
Qulacs & \checkmark & \checkmark & & & & & & & & \\ \hline
\end{tabular}
\caption{Optimization methods: Vector/Matrix partitioning and parallelization (PAR), SIMD vectorization (SIMD), Matrix decomposition (DECOMP), Thread-level parallelism (TRDLVL), 
Linear-algebraic library (LIB), Precision reduction (REDUC), Gate fusion and coalescing (FUS), Tensor flow library  (TENS), Kraus representation  (KRAUS), Qubit recycling (RECYC)}
\label{fig:schOpt}
\end{figure*}

\noindent{\bf Quantum++} \cite{gheorghiu2018quantum}: 
Quantum++ (originally released in 2013) is a computing library in C++17 which contains a quantum computing simulator. The main objectives include the ease of use, high portability, and high performance. Quantum++ models quantum gates in terms of complex vectors and matrices. It optimizes the complexity of computations using parallelism in OpenMP and the use of a linear algebra library (called Eigen3) in a transparent way.  Quantum++ can scale up computations depending on the amount of available memory. For example, it can simulate up to 24 qubits for QFT and 14 qubits for partial trace on the platform that developers use for their experimental evaluations. They use QFT and partial trace as two algorithms for benchmarking and comparison with the performance of Qiskit and QuTiP.

\noindent{\bf LIQUi$\mid >$} \cite{wecker2014liqui}: LIQUi$\mid >$  is an architecture/framework for quantum computing that contains a programming language, optimization and scheduling algorithms, and a multi-platform high performance simulator. It can translate a quantum algorithm written in the form of a high-level program into the low-level machine instructions for a quantum machine. LIQUi$\mid >$  can simulate Hamiltonians, quantum circuits, quantum stabilizer circuits, and quantum noise models. The main abstractions for quantum computation include quantum state vectors and unitary matrices.  LIQUi$\mid >$  enables a modular and extensible design, where users can create their own quantum gates and circuits as reusable and hierarchical components. To minimize the computational cost of simulations,  LIQUi$\mid >$  combines matrices in order to coalesce multiple operations into one. It also provides domain-specific optimization. For example, in quantum chemistry, LIQUi$\mid >$  incorporates domain knowledge about the number of spin ups and spin downs in order to eliminate the cases where they change. The experimental evaluation of LIQUi$\mid >$  is done for Shor's algorithm up to 30 qubits on a desktop machine.  LIQUi$\mid >$  uses 27 qubits to factor a 13-bit number in 5 days (using half a million gates). Quantum programs/algorithms are specified in F$\#$ and then translated into circuit data structures that can later be optimized.

\noindent{\bf NVIDIA’s cuQuantum} \cite{fangNVIDIA2022}: 
NVIDIA cuQuantum SDK provides a high performance library for simulating quantum computations on GPUs. cuQuanum includes two main libraries, namely cuStateVec and cuTensorVec. cuStateVec library provides high performance solutions for state vector-based simulation on classic machines, where quantum states are represented as state vectors of 64/128-bit complex values. Quantum gates are captured as matrices of complex values. The cuTensorNet library provides efficient methods for the contraction of sparse matrices through contraction of high rank tensors to low rank tensors, thereby enabling high performance tensor network computations. The core of optimization methods in cuQuantum includes such tensor network computations on high performance GPUs. As of the time of writing of this paper, no benchmarking data is available on how cuQuantum performs on classic algorithms in terms of circuit width (i.e., the number of qubits).

\noindent{\bf Quantum Circuit Simulator (QCS)} \cite{qCircuit2018}: 
Implemented in Javascript, quantum-circuit smoothly runs simulations involving 25+ qubits within a browser, server, or Jupyter notebook. The simulator supports imports and exports from/to the languages OpenQASM and Quil and exports to pyQuil, Qiskit, Cirq, TensorFlow Quantum, Q\#, and QuEST. As a companion project, Quantum Programming Studio (QPS) is an IDE for this simulator that is also web based. QPS supports visual-based programming with auto generation and drag-and-drop interfacing. This IDE also supports the ability to run code from the GUI to a quantum computer. The entire state vector is stored as a $2^n$ array within a map data structure containing only non-zero amplitudes. Memory usage is no worse than two $2^n$ state vectors as the transformation matrix is not stored. As elements of the transformation matrix are needed, they are generated and applied to the state. The developers of this simulator evaluate it using the following benchmarks: Bell state on the first and the last qubit, Hadamard gates on all qubits, and QFT.



\noindent{\bf Quantum-Kit (Q-Kit)} \cite{tankasalaQuantum-Kit2019}: 
Q-Kit is a closed-source, hybrid quantum-classical circuit simulator designed to run on consumer level computer hardware while still allowing for large simulations. With support for simple GUI based visualizations for design (e.g. the circuit) and analysis (e.g. Bloch spheres and amplitudes), the simulator boasts ease of use.
To support more machines, Q-Kit has a performance mode and a memory mode although Q-Kit operates on only a single CPU core in both instances. Two notable optimizations include the fast translation of measured qubits to classical bits (useful for iterative algorithms, e.g. "recycling" a qubit), and the use of classical computing controls and operations in place of quantum operations. The authors noted that the hybrid quantum-classical circuit design created a large speedup in both memory usage and CPU time. To evaluate the simulator, Shor’s factorization of a 15-bit number using Kitaev’s trick was run. Kitaev's trick refers to the replacement of naive modular exponentiation with quantum phase estimation \cite{kitaev1995quantum}.
More specifically, the exponentiation step is performed by QFT followed by a measurement.
This replacement reduces the number of qubits required for modular exponentiation from linear (relative to the exponent) to logarithmic. Shor's with Kitaev's trick was simulated by Q-Kit using 11 qubits and 32KB of RAM on a consumer desktop computer. For larger simulations, 24 qubits was possible using 8GB while 25+ qubits would require compute servers.


\noindent{\bf QForte} \cite{stair2022qforte}:
QForte is a Python quantum algorithms library for molecular electronic structure simulation.
Included is a general statevector simulator that works with a classical representation of the full quantum state of size $2^n$.
Although the simulator is interfaced using Python, the simulator components are implemented in C++ for efficiency.
Using the Fock space basis, each qubit is stored as a single bit within a 64-bit unsigned integer which allows for efficient bitwise operations.
Moreover, quantum gates are implemented individually with optimization in mind.
For example, the X gate is implemented using low-level C++ copy operations.
QForte also has representations for Hamiltonians for us in the domain-specific quantum electronic structure simulations.
For small, general simulations of about 4 to 12 qubits, QForte is faster than both Cirq and qsim. 
These three simulators perform similarly for medium simulations of 14 to 16 qubits, but QForte is slower for large 20+ qubit simulations, up to 10 times against qsim.

\noindent{\bf Statevector} (IBM Quantum, Qiskit) \cite{IBMQuantum2021}: 
Instead of returning a sampling, the Statevector simulator computes the wavefunction of the statevector ultimately returning the size $2^n$ statevector. This process of wavefunction computation happens continuously as gates and instructions are applied. Statevector supports noisy and ideal simulation with support for up to 32 qubits. 

\noindent{\bf State vector simulator (SV1)} (Amazon Braket) \cite{AmazonBraket, AmazonBraketDevices}: 
The state vector simulator (SV1) is a general-purpose simulator for ideal circuits using any gates without noise modeling.
Locally, there is support for 25 qubits and 34 qubits when fully managed.
Amazon Braket's Tensor simulator is available for a higher number of qubits.
With SV1, the runtime increases linearly with the number of gates; however, this simulator must calculate all possibilities making the runtime exponential in the number of qubits needed.


\noindent{\bf Density matrix simulator (DM1)} (Amazon Braket) \cite{AmazonBraket, AmazonBraketDevices, AmazonBraketExamples}: 
Also known as DM1, it's a simulator for common noisy circuits built using gate noise operations such as bit-flip and depolarizing error. The density matrix is used to describe the pure state state vector in a noisy environment by making a classical mixture of a series of pure states that could result due to the noise.
A quantum channel describes the time evolution of the density matrix by way of the Kraus representation.
Circuits without noise can have noise operations applied to specific gates and qubits to allow the circuit to easily run with noise.
There is support for 12 qubits locally and 17 qubits fully managed.

\noindent{\bf Analog Hamiltonian simulator} (Amazon Braket) \cite{AmazonBraket, AmazonBraketDevices}: 
The Analog Hamiltonian simulator (AHS) simulator supports Hamiltonian programs with one uniform driving field, one (non-uniform) shifting field, and arbitrary atom arrangements. A Hamiltonian is different from a quantum circuit in that gates and qubits are no longer parameters but rather time and space dependent parameters of atoms. The simulator supports 10-12 atoms when run locally.

\noindent{\bf qsim} (TensorFlow Quantum) \cite{broughtonTensorFlowQuantum2020, QuantumAIqsimandqsimh, quantumlibqsimandqsimh}: Designed to run on a single modest machine, qsim is a C++ state-vector simulator that provides the simulation output as a full state vector. With all $2^n$ amplitudes ($n$ is the number of qubits) of the full state vector, sampling comes at little additional computation. With only ~16GB of RAM, qsim can simulate 30 qubits; however, the RAM usage is doubled with each additional qubit. The simulator rather simply performs repeated matrix-vector multiplication, one operation for each gate. The runtime is therefore $O(g\cdot 2^n)$ where $g$ is the number of 2-qubit gates. Several optimizations are applied to decrease the expected runtime: gate fusion, single precision arithmetic, hardware-supported CPU vector instructions (AVX or FMA), and multi-threading via OpenMP. qsim has been used in cross-entropy benchmarking to verify the quantum Sycamore processor \cite{arute2019quantum}. Cross-entropy benchmarking compares how often each bitstring is experimentally observed compared against the results from a simulation.

\noindent{\bf Fock backend} (Strawberry Fields) \cite{killoranStrawberryFields2019, BromleyStrawberryFieldsApplications2020, StrawberryFieldsDocumentation, StrawberryFieldsGitHub}: As a photonic quantum computer simualtor, the Fock backend provides squeezing (noise reduction) and beamsplitter (split and combine light) operations in the Fock basis. For performance, the calculation is made highly vectorized. The simulator encodes quantum computation in a countably infinite-dimensional Hilbert space also in the Fock basis. The infinite dimensions are estimated by imposing a cutoff dimension chosen by the user that limits the number of dimensions used in computation. Unlike the Guassian backend, representation of mixed states are more expensive than pure states and so the pure state is only used in computation when necessary. The discrete Fock states can be created by expanding Gaussian states.

\noindent{\bf Guassian backend} (Strawberry Fields) \cite{killoranStrawberryFields2019, BromleyStrawberryFieldsApplications2020, StrawberryFieldsDocumentation, StrawberryFieldsGitHub}: The Guassian backend uses the Guassian formalism to represent continuous-variable quantum systems in a continuous Guassian state for quantum optical circuits.
As a result, the Guassian backend can support Gaussian Boson Sampling.
Each Guassian state corresponds to a Gaussian distribution.
This backend can calculate the fidelity and Fock state probabilities.
The simulator was written using NumPy.
The computation and representation are more expensive for pure states than mixed states as mixed states are represented via density matrices.

\noindent{\bf TensorFlow backend} (Strawberry Fields) \cite{killoranStrawberryFields2019, BromleyStrawberryFieldsApplications2020, StrawberryFieldsDocumentation, StrawberryFieldsGitHub}: As an extension of the Fock backend, the TensorFlow backend supports TensorFlow tools that allow for optimization and machine learning. The syntax is provided by Strawberry Fields such that the TensorFlow integration is seamless. TensorFlow's automatic differentiation engine is used to train and optimize the provided complex circuits. This integration also supports hardware platforms such as the Xanadu X8 quantum photonic processor.

\noindent{\bf Bosonic backend} (Strawberry Fields) \cite{killoranStrawberryFields2019, BromleyStrawberryFieldsApplications2020, StrawberryFieldsDocumentation, StrawberryFieldsGitHub}: Implemented in NumPy, this backend simulates quantum optical circuits. The simulator represents states as linear combinations of Gaussian functions in phase space. The Bosonic backend is a generalization of the Gaussian backend. The transformations available in the Bosonic backend modify each Gaussian within the linear combination as well as the linear combination's coefficients. All Gaussian, Gottesman-Kitaev-Preskill, cat and Fock states are representable.

\noindent{\bf State Vector} (HybridQ, NASA) \cite{HybridQ}: The HybridQ state vector simulator was designed for high performance computing. Developed in C++, the simulator uses OpenMP for parallelization and JAX for GPU and TPU support. HybridQ was built for AVX instructions whereby both the real and imaginary components of state vectors are stored as contiguous arrays of AVX packed floating point numbers. HybridQ is evaluated against the QuTiP simulator on a random density matrix circuit.

\noindent{\bf cirq.Simulator} (Cirq, Google) \cite{cirq_developers_2022_7465577, CirqDocumentation}: cirq.Simulator is a NumPy sparse matrix state vector simulator for pure states. The simulator can mimic actual quantum hardware such that, for example, the initial state is the all zeros state and the state vector is inaccessible. Runs can even be stochastic. The simulator, however, has the ability to give the state vector at the end of the simulation and the initial state can take a full state vector. cirq.Simulator supports pure and noisy simulations. Noisy simulations use Monte Carlo wavefunction simulation whereby a Kraus operator is randomly sampled and applied to the wavefunction.

\noindent{\bf cirq.DensityMatrixSimulator} (Cirq, Google) \cite{cirq_developers_2022_7465577, CirqDocumentation}: This density matrix simulator is Cirq's best for simulating noisy quantum circuits and mixed states. The simulator allows the user to view the density matrix as the simulation steps through the circuit. The simulator may optionally mimic quantum hardware. The density matrix is accessible at the end of the run, but the result of any measurements will appear in the density matrix. The initial state for a simulation can be a full density matrix or a full wave function.

\noindent{\bf QuIDE} \cite{QuIDESim}:
Quantum IDE (QuIDE) is a simulation platform for the usage of students in understanding large quantum circuits.
The Microsoft .Net Framework was used to develop the QuIDE core simulation library, QuIDE.dll, which is a standalone C\# library. Internally quantum states are stored in a collection of quantum register classes. Quantum register classes are built of a \texttt{<key, value>} dictionary with the keys representing a single state of the possible quantum states and a value representing the probability of that state given the whole quantum state. Composed quantum gates are applied across these dictionaries by decomposing their operations to basic gates such as Not, CNot, and matrix-defined single qubit gates. The implementation of QuIDE features a novel approach to creating quantum algorithms as at any point a coded representation can be converted to its graphical representation and vice versa. QuIDE also provides two solutions for the simulation of quantum algorithms :a black box approach as would be observed in a true quantum computer, and a step by step approach that is better suited to understanding quantum computations. Using QuIDE, the researchers were able to demonstrate the ability to use 23 qubits in a quantum algorithm as well as factoring up to 12 bits of a number using Shor's algorithm. The researchers strove to provide a novel teaching tool that increased the efficiency of understanding quantum computing. Usability evaluation tests performed by the researchers scored QuIDE higher than similarly powerful simulators within the field. These conclusions are supported by the interchangeable approach QuIDE takes to creating and displaying data.

\noindent{\bf Quantum Computing Playground} \cite{QCPlayground}: Quantum Computing Playground is a simulator and IDE for experimenting with graphical acceleration of large quantum algorithms. Quantum Computing Playground was developed by a group of Google engineers and is an open source web application that simulates quantum computers. Development was done in JavaScript and intended for standalone web deployment. Scripts and algorithms created in this IDE's proprietary QScript language are all public and accessible through \url{https://www.quantumplayground.net/}. Quantum Computing Playground provides a unique approach to GPU acceleration by interpreting quantum states as textures and processing gates by using GLSL shaders. The internal representations of an algorithm's quantum states are stored in textures; For instance a quantum algorithm containing 6 qubits would have 64 unique quantum states which would be represented by a $8\times 8$ texture with the red and green components determining the real and imaginary components of each state. Quantum gates are processed internally on a single qubit at a time using a unique pixel based GLSL shader within the WebGL library. Due to the nature of shader computations, the calculations are quick and are applied to all qubits within a texture, but may only affect a portion of the texture. These textures are currently limited to a height and width of 2048 pixels, thereby limiting this simulator to 22 qubit calculations. This texture-based approach mimics other matrix-combination-based solutions, yet it is novel in the storage of quantum states as well as the optimization techniques. The intent behind Quantum Computing Playground is to experiment with graphical acceleration and provide three dimensional visualizations for quantum computations and algorithms.
As this is the scope of development, no benchmark data is provided.
Scripts for major algorithms such as Grover's algorithm and Shor's factoring algorithm are provided.

\noindent{\bf ProjectQ} \cite{Steiger2018projectqopensource}: ProjectQ is an open-source software framework for quantum computing started at ETH Zurich, with developers branching out to work for companies such as Microsoft and Huawei. ProjectQ is a python framework that provides users with a quantum circuit based simulator as well as a quantum emulator. Of the two, the Quantum emulator is the true reason for utilizing this framework. Internally, quantum registers are represented by arrays and quantum gates are decomposed into minimal matrices and applied across the quantum register. The emulator adds to the simulator by utilizing the advantages of classical computation when encountering certain decompositions, such as QFT decompositions. Instead of performing each Hadamard gate individually, a classical computation is done which turns out to be quicker. ProjectQ also provides support for running Quantum algorithms developed within its framework on true quantum computers such as IBM's 5-qubit quantum computer. The developers seem intent on providing shortcuts to quantum compilation times as well as providing resources for debugging quantum algorithms.

\noindent{\bf Sparse Simulator} (Azure) \cite{Azurequantum}: The Azure sparse simulator is a simulator developed by Microsoft for their Azure Quantum program and Quantum Development Kit, noted in the frameworks section. The sparse simulator is similar to other full state simulators but works on multiple portions of a quantum vector instead of the full quantum vector. In algorithms where much of the full state would be zero, it allows for a larger amount of qubits to be simulated. This is done by only simulating a small portion of the full state. Specifically, only what is non-zero. This implementation allows for algorithms with differing amounts of extraneous computation time to be tested with more or fewer qubits. This simulator, at the moment, can be utilized on a variety of platforms using the Azure tools and Q\# language, but is not accessible otherwise.

\noindent{\bf Toffoli Simulator} (Azure) \cite{Azurequantum}: The Toffoli simulator is a simulator developed by Microsoft for their Azure Quantum program and Quantum Development Kit, noted in the frameworks section. The Toffoli simulator acts identically to full state simulators, but puts limitations on the gates available to algorithms. By limiting the possible gates in a quantum algorithm, it is able to simulate millions of qubits simultaneously. Large scale qubit tests can then be done on a subset of possible quantum algorithms. This simulator, at the moment, can be utilized on a variety of platforms using the Azure tools and Q\# language, but is not accessible otherwise.

\noindent{\bf Forest SDK} \cite{QuilISA}: Forest SDK is developed by Rigetti Computing: a developer of quantum integrated circuits as well as quantum simulators. Development was done using a proprietary quantum instruction set Quil detailed in \cite{QuilISA}. Quil features functionality such as "Applying arbitrary quantum gates, Defining quantum gates as optionally parameterized complex matrices, Defining quantum circuits as sequences of other gates and circuits, which can be bit- and qubit parameterized, Expanding quantum circuits, Measuring qubits and recording the measurements into classical memory, Synchronizing execution of classical and quantum algorithms, Branching unconditionally or conditionally on the value of bits in classical memory, and Including files as modular Quil libraries such as the standard gate library". Internally Quil's QVM or Quantum Virtual Machine represents a quantum state as an array of values, and applies a strict set of quantum gates specified in the quantum instruction set. The intent behind Forest is to support Rigetti Computing's Quantum Abstract Machine (QAM) which is a quantum state machine used for simulating arbitrary quantum algorithms. In effect, Quil causes state changes within QAM. By working directly with instruction sets as well as the quantum integrated circuits, Forest SDK is a cloud based quantum algorithm solution which is able to provide both a powerful tool for describing quantum algorithms as well as pushing the boundaries on quantum simulation efficiency.

%% file: decision.tex
\subsection{Decision Diagram}
\label{sec:decD}

This section discusses methods based on Quantum Information Decision Diagrams (QuIDDs). QuIDDs are data structures in the form of binary Directed Acyclic Graphs (DAGs) that provide a compact method for the storage of quantum states in memory. Each quantum state of dimension $n$ may include up to $2^n$ complex amplitudes whose storage is a challenge. QuIDDs exploit redundancies in the values of amplitudes by using the binary representation of each basis state as a routing direction. For example, let $n=3$ and consider a quantum state $\Phi=\Sigma_{i=0}^{n-1} \alpha_i |x \rangle$, where $|x \rangle$ denotes the Ket notation of some basis state $x$ out of 8 possible states. Suppose we are looking for the amplitude $\alpha_{010}$. Then, starting from the root of the QuIDD that stores $\Phi$, we take the left branch when we see a 0 and take the right branch upon observing a 1. The edges of the DAG are labeled (including the entry edge to the root) by complex values whose multiplication down the path $010$ would give us the amplitude of $\alpha_{010}$. A similar approach can be used to store unitary matrices (i.e., quantum gates). Next, we review some of the most important simulators developed based on QuIDDs.

\noindent{\bf QuIDDPro} \cite{viamontes2003gate,viamontes2004graph,viamontes2007efficient,viamontes2009quantum}: 
QuIDDPro is a high performance, scalable, and easy-to-use quantum circuit simulator for applications such as quantum computation and communication. The core novelty of QuIDDPro is in using QuIDDs to model state vectors and density matrices. The developers of QuIDDPro use clever matrix multiplication techniques in order to avoid explicit creation of the operator (gate) matrices. They optimize quantum operations when matrices have blocks of repeated values. To evaluate the efficiency of QuIDDPro, its developers conduct two classes of experiments based on the density matrix model and the state vector model. Experiments using density matrix model evaluate QuIDDPro using circuit of up to 24 qubits and 21622 gates in reversible logic, quantum communication, quantum error correction, and quantum search. Comparisons with QCSim show the superiority of QuIDDPro. Experiments using the state vector model also show that QuIDDPro outperforms other simulators when evaluated on Grover's search algorithm of up to 40 qubits. Designers can specify quantum circuits using a script language (similar to MatLab).

\noindent{\bf MQT DDSIM} \cite{zulehner2018advanced,burgholzer2021hybrid}:
MQT is a simulator of classical quantum circuits that uses  QuIDD to represent quantum states in memory. The core unit of design includes the decision diagram, which provides abstraction and modularity for representing quantum states and transformations. MQT uses QuIDD to compactly represent quantum states, unitary transformations and their Kronecker product. The QuIDD representation has the potential to avoid exponential space complexity of matrices and vectors of an $n$-qubit system. QuIDD is a totally different way of storing quantum states and transformations with respect to array-based methods (e.g., LiQUi$\mid >$, QX, ProjectQ) where one uses one or two dimensional arrays. Developers of MQT evaluate its performance on QFT, entanglement, Grover's and Shor's algorithms in comparison with Microsoft's LiQUi$\mid >$, QX, ProjectQ, and QuIDDPro. They show that MQT can simulate quantum circuits of up to 100 qubits for entanglement, up to 64 qubits for QFT, up to 40 qubits for Grover's algorithm, and up to 37 qubits for Shor’s algorithm. 

%% file: vqa.tex
\subsection{Variational Quantum Algorithms}
\label{sec:vqa}

Variational Quantum Algorithms (VQA) provide a new paradigm for the creation of quantum circuits for NISQ computers. All quantum machines in use today are NISQ devices which come with considerable constraints in the limitations of qubits in quantity, connectivity, and circuit depth. The creation of quantum circuits that fit within these constraints while retaining efficiency is difficult. As the name implies, NISQ machines are troubled with noise: a constraint left unconsidered by the quantum algorithms designed with the fault-tolerant era in mind.
Using classical machine learning methods, VQA facilitates the creation of NISQ circuits automatically. Such methods include noise mitigation by optimizing for shallow circuit depth. VQA makes use of classical optimizations applied to parameterized quantum circuits. For example, parameter optimization is applied using the same standard classical optimizers that are found in machine learning applications \cite{CerezoVQA2021}.

\noindent{\bf Yao.jl} \cite{luoYao.jl2020}: The Yao simulator has a focus on high performance in medium and smaller sized quantum circuits. As an open-source framework, the simulator includes extensibility as a major objective. Implemented in the Julia programming language \cite{bezanson2017julia}, Yao is made as an extensible Julia library. For this reason, any IDE support will be through Julia. Yao uses differentiable programming such that the quantum circuit is learned by an objective function. This gradient-based approach allows the use of neural network optimizations such as GPU acceleration. Also, Yao adapts to the new problem space with its automatic differentiation engine that offers speed and constant memory cost against circuit depth. Yao uses a tensor representation for inner calculations of quantum states. Yao was evaluated with various gates against the number of qubits: X, H, CNOT, and T. This evaluation tests up to 25 qubits.

\noindent{\bf PennyLane} \cite{bergholmPennyLane2018}: PennyLane is an open-source, Python-based framework for training variational quantum circuits using automatic differentiation as it is a similar problem to machine learning. The ultimate motivation behind PennyLane is to advance both machine learning and quantum computing. PennyLane supports hybrid quantum-classical computation with a focus on quantum machine learning on near-term quantum devices. When the devices are ready, machine learning will then be processable on quantum computers. Tying hybrid computation to differentiable programming means that PennyLane supports the classical steps in a differentiable manner such that both the classical and quantum components are optimized in tandem.
Since PennyLane is a framework, it extends several machine learning libraries such as TensorFlow and PyTorch.
Similarly, it supports easy integration of external quantum devices and provides five internal simulators with different properties with support for at least 20 qubits. Several external quantum devices such as StrawberryFields, IBM's Qiskit, and Google Cirq are already supported.

\noindent{\bf Qulacs} \cite{suzuki2021qulacs}: 
Qulacs is a simulator for fast simulation of large quantum circuits/algorithms. Qulacs applies several optimization methods to improve time efficiency of simulation.  
Qulacs contains three libraries used in a layered approach. The first library, which is also the lowest layer, is written in C and ensures efficient memory management for the representation and updates of quantum states and the unitary transformations applied to them. The second library raises the level of abstraction and provides object-oriented concepts for the creation of quantum gates and circuits. At runtime, the second layer also chooses the most efficient implementation of quantum gates based on the circuit width (i.e., number of qubits). The third library wraps gates and circuits as variational quantum algorithms that can be trained. While Qulacs enables high-performance simulation of large circuits, it also minimizes overhead for redundant execution of small circuits (up to 10 qubits). Qulacs can represent quantum states based on (i) Schrodinger’s approach (i.e., state vector for quantum states and density matrices for quantum gates) due to its time efficiency; (ii) Feynman’s method with less memory requirements, and (iii) Hybrid approach depending on the width and depth of the circuit being simulated. For example, Qulacs uses tensor-network-based simulators for simulating low depth-high qubit circuits. Qulacs utilizes several techniques for optimization such as SIMD vectorization in modern CPUs (e.g., Intel AVX2 up to 256 bit words processed simultaneously; i.e., 4 real values, which is two complex values processed simultaneously), multi-threading with OpenMP, quantum circuit optimizations and GPU-based acceleration. Experimental evaluations show that Qulacs can simulate a variety of randomized circuits with up to 25 qubits (e.g., randomized layered circuits of X-Y-Z rotation in one layer followed by a CNOT layer).

%% file: tnc.tex
\subsection{Tensor Network Contraction}
\label{sec:tnc}

Simulation based on Tensor Network Contraction (TNC) considers a quantum circuit diagram as a network of tensors (a.k.a. circuit graph) and progressively contracts the tensor network until there is only one vertex left. The contraction of two tensors is actually a convolution: the product of a matrix multiplication. 
As such, the final vertex is the sum total of the wave function and represents a sample of the amplitudes of a quantum circuit. In this process, some gates might be partially simulated, thereby providing a potential advantage over other simulation methods. A defining feature of TNC simulators is the optimization of contraction order. Contraction costs are often determined by the nature of the quantum gates within an algorithm.
Large quantum gates result in tensors that create sequential contraction orders while unitary and binary quantum gates create more optimal orderings.
The costs of such contractions can be evaluated by the tree-width of their circuit graph (tensor network). The authors of \cite{markov2008simulating} show that any quantum circuit whose graph has a logarithmic tree-width can be simulated deterministically in polynomial time on classic machines.

\noindent{\bf Tensor network simulator} (Amazon Braket) \cite{AmazonBraket, AmazonBraketDevices}: The Tensor Network simulator (TN1) can simulate certain circuits up to 50 qubits at a circuit depth of up to 1000. Simultaneously, TN1 provides the ability to run many instances of the quantum measurement, and view the outcomes. Out of the available Amazon Braket simulators, TN1 is recommended for circuits with any of the following properties: sparse, local gates, or special structures (e.g. QFT). This simulator has two phases: the rehearsal phase and the contraction phase. The contraction phase is only attempted if the rehearsal phase can successfully identify an efficient computation path that can be executed within Amazon Braket's runtime limits (when running managed).

\noindent{\bf MPS} (IBM Quantum, Qiskit) \cite{IBMQuantum2021, QiskitDocumentation}: Using the Matrix Product State (MPS) representation, this tensor-network simulator is most optimal for quantum states with weak entanglement. There is support for ideal modeling up to 100 qubits. The MPS representation uses a one-dimensional array of tensors to form any quantum state in a Hilbert space with $D$ dimensions. Certain Hamiltonians can be represented in one dimension and efficiently approximated under a finite $D$ with almost arbitrary accuracy. In other cases where higher dimensions are required for the tensor, polynomial or exponential growth can be seen in the required space \cite{OrusMPS}. For simulations, the tensor remains small for circuits with many single-qubit gates and few two-qubit gates \cite{QiskitDocumentation}.

\noindent{\bf Quimb} \cite{gray2018quimb}: Quimb is a specialized Python library crafted for the efficient simulation and analysis of quantum many-body systems. In the context of this paper, we can think of quantum many-body systems as systems composed of quantum particles capable of interacting and entangling with one another. In essence, Quimb performs similar computations as other quantum computer simulators. It achieves this by utilizing Python libraries like NumPy to facilitate the representation and computation of these quantum many-body systems. Quimb excels in handling swift tensor network contractions, which can involve anywhere from 100 to 1000 tensors. These tensors can be interconnected in a variety of ways, including shapes such as matrix product states or more intricate configurations. The limiting scalability factor for this simulator, as with other tensor network contraction simulators, is the creation of an optimal contraction path. With its powerful capabilities, Quimb empowers researchers, enabling them to gain valuable insights into the behavior of complex quantum systems with a simple and efficient process.

%% file: feynman.tex
\section{Feynman Method}
\label{sec:feynman}

This section discusses the Feynman path integrals method for the simulation of quantum circuits.
Although no simulators are listed under the Feynman method, research is ongoing.
We predict that this ongoing research will lead to new simulators and improvements to existing simulators.
Therefore, we present Feynman simulation algorithms instead for completeness.

\noindent{\bf Feynman Method}
In an $n$-qubit complex Hilbert space, the probability amplitude of transiting from one basis state to another is specified as the summation of the probability amplitudes of all possible paths to that state. For example, when $n=1$, one can think of a Hadamard gate applied to a single qubit $\ket{x}$ as a transformation that generates the following state: $\ket{\phi} = (1/\sqrt2) \Sigma_{y \in \{0,1\}} (-1)^{xy} \ket{y} = (1/\sqrt2) (\ket{0} + (-1)^{x} \ket{1})$. This state is the summation of the probability amplitudes of two possible states $\ket{0}$ and $\ket{1}$ depending on the input state $\ket{x}$. The variable $y$ is called the {\it path variable} because its values represent the computational paths simultaneously taken by the H gate. 
Thus, the summation $\ket{\phi}$ is called a {\it sum-over-paths} or a {\it path integral} (for two paths $x \mapsto \ket{0}$ and $x \mapsto \ket{1}$). 





Generalizing this idea to an $n$-qubit quantum circuit, a {\it Feynman path} captures a sequence of basis states from an input state $\ket{x_{n-1}, x_{n-2}, \cdots, x_{0}}$ to a specific output state $\ket{y_{n-1}, y_{n-2}, \cdots, y_{0}}$ through a set of intermediate states $\ket{z_{n-1}, z_{n-2}, \cdots, z_{0}}$.
The state of each individual qubit $b_i$ in $\ket{x}$ (where $1 \leq i \leq n$) changes through a sequence of intermediate transformations. It is known that such transformations change a qubit only when the gates are $H, X^{1/2}, Y^{1/2}$, where $H= \frac{1}{\sqrt 2}$
$\begin{pmatrix}
1 & 1\\
1 & -1
\end{pmatrix}$, $X^{1/2} = 1/2$
$\begin{pmatrix}
1+i & 1-i\\
1-i & 1+i
\end{pmatrix}$, and $Y^{1/2} = 1/2$
$\begin{pmatrix}
1+i & -(1+i)\\
1+i & 1+i
\end{pmatrix}$ for $X = $
$\begin{pmatrix}
0 & 1\\
1 & 0
\end{pmatrix}$ and $Y = $
$\begin{pmatrix}
0 & -i\\
i & 0
\end{pmatrix}$.  Notice that, $H, X^{1/2}, Y^{1/2}$ have two non-zero values in both rows and columns; i.e., non-diagonal gates. This means that the value of a qubit flips only when non-diagonal gates are applied to it. 
The sequence of values of $b_i$ determines the value of $b_i$ after the $k$-th step, for $1 \leq k \leq m$, in  a circuit  of depth $m$. The output of a sequence of $m$ unitary gates applied on $\ket{x}$ can be obtained through computing $mn$ binary values; i.e., {\it path integrals}. As a result, when simulating the computations of a circuit $C$ on an input $\ket{x_{n-1}, x_{n-2}, \cdots, x_{0}}$, one can compute the path integrals for an expected output $\ket{y_{n-1}, y_{n-2}, \cdots, y_{0}}$ without a need for calculating all possible amplitudes in the output quantum state/superposition. 


\noindent{\bf Stabilizer}.
Huang and Love \cite{HuangFeynmanStabilizer2021} propose a Feynman simulation method and a Hybrid Schr\"odiner-Feynman simulation method called stabilizer-based path integral recursion (SPIR) and stabilizer projector contraction (SPC) respectively.
We explore the former. The SPIR algorithm modifies the Feynman path integral by dividing a circuit into layers of Clifford and non-Clifford gates where a Clifford layer contains only the gates of types H, CNOT, and the $\pi/4$ phase gate S.
The Clifford group of gates map Pauli matrices I, X, Y, and Z to themselves.
The Clifford layers can be captured by Feynman path integrals, and the part of the circuit with non-Clifford layers is recursively divided into two sub-circuits.
Each unitary transformation $U$ in the non-Clifford layers can be captured by the  decomposition $U = \Sigma^k_{i=1} c_i \ket{\phi_i} \bra{\phi_i}$, called the {\it stabilizer projectors}.
At the base case of the recursion, $2^k$ inner-products are calculated, for a cost of $O(n^32^k)$ for each layer.
The final amplitude is built as the recursion returns values.


The non-Clifford layers are rewritten into sums of stabilizer projectors.
The Feynman path method has a time cost of $O(4^m)$ and a memory cost of $O(m + n)$ where $m$ is the number of gates and $n$ is the number of qubits.
The SPIR method has a time cost of $O(n^3(2\cdot d_{nc})^k)$ and memory cost of $O(n \log{d_{nc}})$ where $d_{nc}$ is the depth of non-Clifford layers and $k$ comes from the stabilizer projector rank $2^k$, where $k \thicksim n$, for each layer of gates.
In this way, the time cost of this approach is in terms of the number of qubits and circuit depth rather than the number of gates.
Although the circuit depth relates to the number of gates, this is still a significant time advantage for many circuits while requiring not much more memory.
The SPIR method relies on the fact that layers of stabilizer projectors can be used to replace non-Clifford gates.
Subsequent operations involve taking the inner product between stabilizer states, which takes $O(n^3)$ time, scaling by the number of terms in the stabilizer projector decomposition of each gate.

\noindent{\bf Feynman-based Simulators}.\ At the time of this writing, we have found no simulator that functions  solely based on the Feynman path integrals methods. The only noteworthy tool we would like to mention here is the verifier FEYNMAN \cite{AmyQPL2018,amy2019formal} that implements a functional verification method for quantum programs based on the Feynman path integrals. We note that this is an area in need of further investigations.

%% file: hybrid.tex
\section{Hybrid Schr{\"o}dinger-Feynman Method}
\label{sec:hybrid}

This section presents simulation methods that utilize both Schr{\"o}dinger-based and Feynman's path integral approaches for improving the scalability of simulations, called the Hybrid Schr\"odinger-Feynman methods 
\cite{markov2018quantum,arute2019quantum}. The most important simulators in this category include the DDSIM \cite{burgholzer2021hybrid}, Rollright \cite{markov2018quantum}, Jet \cite{Vincent2022jetfastquantum}, and qsimh \cite{broughtonTensorFlowQuantum2020, QuantumAIqsimandqsimh, quantumlibqsimandqsimh}.

\noindent{\bf DDSIM}.\ Burgholzer {\it et al.}  \cite{burgholzer2021hybrid} present a hybrid Schr\"odinger-Feynman method for the simulation of quantum computations using decision diagrams. To capture quantum transformations by QuIDDs, matrices are recursively broken down into $2\times 2$ matrices. Each node of the QuIDD represents a $2\times 2$ matrix at some level of recursion, starting from the root that captures the highest level. Each node has four children where each outgoing edge captures the amplitude of one of the sub-matrices from left to right. For example, for the matrix $U=\begin{pmatrix}  U_{00} & U_{01}\\   U_{10} & U_{11}\end{pmatrix}$ the first node of its QuIDD has four children with indices 00, 01, 10, and 11 from left to right. The weights of these arcs represent the amplitudes multiplied by each of the four sub-matrices. Thus, the application of a unitary on a quantum state would be captured as the multiplication of two QuIDDs. If the multiplication of the two QuIDDs remains compact then simulation of quantum computation can be done efficiently. However, this may not be the case always. To tackle this issue, the authors of \cite{burgholzer2021hybrid} present a method that combines QuIDD-based simulation with Feynman’s paths integral approach. They apply Schmidt decomposition to 2-qubit gates (i.e., $4 \times 4$ unitaries)  represented as the summation of four tensor products $\Sigma (\ketbra{i}{j}) \otimes U_{ij}$ for $i, j \in \{0, 1\}$. Each term in this sum can be computed using independent QuIDDs and then summed up. The proposed hybrid approach creates horizontal cuts in a given circuit to split the input qubits into separate groups. The gates that are across such cuts are decomposed (using Schmidt’s approach) and simulated separately. Such cross-block gates determine the number of independent simulation runs. For example, two cross-block gates that are connected to each other would require 4 independent simulation instances. Thus, the number of simulation runs grows exponentially in terms of the number of cross-blocks. As such, this approach is suitable for circuits with a limited depth.

\begin{figure*}[h]
\centering
\begin{tabular}{ |p{18mm}||p{7mm}|p{9mm}|p{16mm}|p{14mm}|p{7mm}|p{13mm}|p{6mm}|p{9mm}|p{13mm}|p{13mm}| }
 \hline
 \multicolumn{11}{|c|}{{\bf Optimization Techniques in Hybrid Schr{\"o}dinger-Feynman-based Simulators}} \\ \hline
 {\bf Simulator} & {\bf PAR} & {\bf SIMD} & {\bf DECOMP}  &  {\bf TRDLVL}  &  {\bf LIB}  &  {\bf REDUC}  &  {\bf FUS}  &  {\bf TENS} &  {\bf KRAUS} & {\bf RECYC} \\ \hline
DDSIM & & & \checkmark & & & & & & & \\ \hline
Rollright & \checkmark & & \checkmark & \checkmark & & & \checkmark & & & \\ \hline
qsimh & \checkmark & & & \checkmark & & & & & & \\ \hline
Jet & \checkmark & \checkmark & \checkmark & \checkmark & & \checkmark & & & & \\ \hline
Tensor network simulator & & & & & & & \checkmark & & & \\ \hline
MPS & & & & \checkmark & & & \checkmark & & & \\ \hline
Tensor Contraction & & \checkmark & & \checkmark & & \checkmark & \checkmark & & & \\ \hline
Quimb & & & & \checkmark & & & \checkmark & & & \\ \hline
qFlex & \checkmark & & & & \checkmark & \checkmark & & & & \\ \hline
\end{tabular}
\caption{Optimization methods: Vector/Matrix partitioning and parallelization (PAR), SIMD vectorization (SIMD), Matrix decomposition (DECOMP), Thread-level parallelism (TRDLVL), 
Linear-algebraic library (LIB), Precision reduction (REDUC), Gate fusion and coalescing (FUS), Tensor flow library  (TENS), Kraus representation  (KRAUS), Qubit recycling (RECYC)}
\label{fig:hybOpt}
\end{figure*}

\noindent{\bf Rollright} (University of Michigan, Google) \cite{markov2018quantum} Rollright is an innovative hybrid Schrödinger-Feynman simulator designed for quantum computing tasks, eliminating the need for Inter-Process Communication (IPC) or proprietary hardware. This simulator consistently outperforms its counterparts like Qiskit and Microsoft's QDK in both scalability and computational speed. The key optimizations of Rollright occur within its clustering and gate decomposition stages. During clustering, qubits in close proximity are grouped based on their gate usage, and controlled gates are placed on cluster borders for easy decomposition. Schrödinger-style gate calculations are performed on each cluster to determine its result, which is later integrated into the overall computation. In the decomposition stage, controlled gates on cluster edges are decomposed into matrices, effectively separating these clusters. The paper highlights cross-cluster controlled Z gates, mirroring Feynman path integral relationships. These stages can be adjusted for trade-offs between accuracy and size, providing flexibility in handling quantum computing tasks efficiently.


\noindent{\bf qsimh} (TensorFlow Quantum, Google) \cite{broughtonTensorFlowQuantum2020, QuantumAIqsimandqsimh, quantumlibqsimandqsimh}: qsimh is a hybrid Schrödinger-Feynman simulator extension for qsim. The simulator is built within the same library as qsim but with modifications to become a hybrid Schr\"odinger-Feynman simulator. This extension to qsimh supports Feynman path integral simulation through the quantum algorithms lattice being split into two parts and the Schmidt decomposition being used to decompose 2-qubit gates on the cut. As each path is independent, parallelization across single machines and multiple machines are supported. The full state vector is not computed by a single task but the combination of many. Moreover, it is possible to retrun selected amplitudes instead of the full output vector. Despite the parallelization abilities, the resource consumption of qsimh still grows exponentially with respect to the number of qubits, and the number of paths. However, qsimh boasts support for 50+ qubits.

\subsection{Tensor Network Contraction}

Expanding on TNC, the following simulators may utilize different methodologies in order to slice the contraction space.
The slicing of a tensor network contractions path can be viewed as an interpolation between the methodologies of a Schr\"odinger simulator and a Feynman simulator \cite{Huang2021}.

\noindent{\bf Jet} \cite{Vincent2022jetfastquantum}: Jet is an open source tensor network simulator that is a cross-platform C++ and Python library.
Jet aims to exploit task-based parallel processing and significant workload overlap during the contraction of slices of tensor networks.
In order to address the changing landscape of super computers as well as drawbacks of current memory technology, a new methodology for separating tensor network contractions into slices was developed.
Slices work to compact memory requirements by setting a variable within the tensor network equation to either 0 or 1 and computing the contraction \cite{Huang2021}.
The result of such slices is an incomplete wave form tensor that can be completed by computing the respective opposite slice. For instance, if the first slice had a variable $x$ set to 0 to complete the tensor, then one would calculate the slice where $x$ is equivalent to 1.
This operation works to reduce tensor size by allowing for all the tensor nodes to be removed from one part of the given simulation.
By generating a task-based dependency graph and utilizing slicing, Jet is able to easily parallelize TNCs through task-based parallelism frameworks such as Taskflow Library \cite{9200667}.
Jet then provides one more optimization by computing tensor contractions that will occur across many slices and removes the redundant calculation by storing the results of these contractions in memory.
Jet is able to efficiently simulate up to 53 qubits with a circuit depth of 10 through slicing; however, system memory constraints may limit full circuit simulation to about 27 qubits.
Several configurations of the Sycamore-53 and Gaussian Boson Sampling (GBS) supremacy circuits are used as benchmarks.
The Sycamore-53 circuit is a pseudo-random circuit used by Google in 2019 to demonstrate quantum supremacy \cite{arute2019quantum}.
The GBS circuit is a variation of the Bosom Sampling problem in photonic computing.
The Gaussian variation considers a general Gaussian state in the Fock basis \cite{kruse2019gbs, deshpande2022gbs, hamilton2017gbs}.
At a depth of 10, Jet is able to solve a single amplitude output in 0.76 seconds; estimates for a single amplitude at a depth of 20 are around 35 seconds \cite{Vincent2022jetfastquantum}.
These estimates should put a full simulation of the task at around 58 years.
Other benchmark algorithms such as Shor’s algorithm or Grover’s algorithm do not have definitive testing available, but it is clear to see, given the performance of the simulator and its relative closeness to the Sycamore Circuit, such benchmarks should perform well.

\noindent{\bf Tensor Contraction} (HybridQ, NASA) \cite{HybridQ}:
CoTenGra uses HybridQ to identify the best contraction and then executes the contraction using Quimb. CoTenGra applies tensor slicing when appropriate. It supports multi-threading through OpenMP or MKL, and parallelizing the contraction search is also possible. The Message Passing Interface (MPI) is supported to enable the distribution of work among nodes, with a final recollection occurring through a divide-and-conquer approach. The evaluation of HybridQ was conducted on a density matrix circuit of 50 gates.

\noindent{\bf qFlex} \cite{villalonga2019qflex}: qFlex, a cutting-edge tensor network contraction simulator, jointly developed by Google and NASA, stands out for its exceptional versatility and speed compared to other simulators in its category. Remarkably, it maintains this performance without compromising its ability to handle arbitrary quantum circuit. Two key optimizations underpin its efficiency: firstly, it leverages the tensor contraction of CZ gates, condensing the eight layers around each CZ gate into a much smaller matrix. Secondly, it is able to divide large tensor contractions using scalars and then processing the chunks. The culmination of these advancements makes qFlex one of the top-performing quantum simulators available today, even serving as a benchmark for quantum supremacy experiments.

%% file: heisenberg.tex
\section{Heisenberg Method}
\label{sec:heisen}
This section presents an overview of Heisenberg simulation method. We first provide the core idea behind the Heisenberg approach, and then Subsection \ref{sec:stab} discusses stabilizer circuits.

Schr{\"o}dinger-based simulation of quantum computations applies unitary transformations to quantum states, which results in the evolution of quantum state. To classically calculate such state evolution for $n$ qubits, one needs to perform the multiplication of a $2^n \times 2^n$ matrix $U$ of complex values by a $2^n \times 1$ vector $|\phi \rangle$ of complex values. Matrix $U$ represents the quantum circuit $C$ that takes $n$ input qubits. Once the output state is computed, measurement of $d \leq n$ qubits might be done to find the solution of a problem. In the case of a decision problem, we just need to measure one output qubit, where the value of the qubit would determine the answer to the decision problem. Thus, we consider one additional answer bit. The initial input to $C$ would be $\ket{0 0 \cdots 0 0}$ as an $n+1$-qubit string. This way, we have two outcomes  $\ket{0 0 \cdots 0 0}$ and  $\ket{0 0 \cdots 0 1}$ depending on the answer of the decision problem. Each measurement outcome has its own probability proportional to the square of the magnitude of the complex amplitudes of $\ket{0 0 \cdots 0 0}$ and  $\ket{00 \cdots 0 1}$. Trivially, the time and space costs of performing such matrix multiplications is exponential in $n$, but we will only have access to the outcome of a measurement with some probability without being able to know all the complex values involved in state evolution. What if one could just determine the measurement probabilities without actually computing $U\ket{\phi }$? Can such computation of measurement probabilities be done in polynomial time on classic machines? The answer depends on the nature of $C$, and such a simulation method is called a {\it stateless} simulation.
In summary, the Heisenberg simulation method is a stateless approach that can determine the outcome probabilities in polynomial time on classic machines if $C$ is a stabilizer circuit. 


\begin{figure*}[h]
\centering
\begin{tabular}{ |p{18mm}||p{7mm}|p{9mm}|p{16mm}|p{14mm}|p{7mm}|p{13mm}|p{6mm}|p{9mm}|p{13mm}|p{13mm}| }
 \hline
 \multicolumn{11}{|c|}{{\bf Optimization Techniques in Heisenberg-based Simulators}} \\ \hline
 {\bf Simulator} & {\bf PAR} & {\bf SIMD} & {\bf DECOMP}  &  {\bf TRDLVL}  &  {\bf LIB}  &  {\bf REDUC}  &  {\bf FUS}  &  {\bf TENS} &  {\bf KRAUS} & {\bf RECYC} \\ \hline
Stabilizer and Extended Stabilizer & & & \checkmark & & & & & & & \\ \hline
CHP & & & & & & & & & & \\ \hline
cirq.CliffordSimulator & & & & & & & & & & \\ \hline
Stim & & \checkmark & & & & & & & & \\ \hline
Clifford Expansion & & & & \checkmark & & & \checkmark & & & \\ \hline
\end{tabular}
\caption{Optimization methods: Vector/Matrix partitioning and parallelization (PAR), SIMD vectorization (SIMD), Matrix decomposition (DECOMP), Thread-level parallelism (TRDLVL), 
Linear-algebraic library (LIB), Precision reduction (REDUC), Gate fusion and coalescing (FUS), Tensor flow library  (TENS), Kraus representation  (KRAUS), Qubit recycling (RECYC)}
\label{fig:heiOpt}
\end{figure*}



\subsection{Stabilizer}
\label{sec:stab}
Clifford circuits solely use Clifford gates, a subset of quantum gates, and are therefore not universal circuits. A Clifford circuit requires $H$, $S$, and either $CX$ or $CZ$ as the minimally generating set. By the Gottesman-Knill theorem \cite{GottesmanPhD}, Clifford circuits can be simulated in polynomial time in the number of gates and qubits. There exist several different Clifford circuit simulation algorithms. This efficiency is due to the fact that, for a circuit $C$ on input stabilizer states (such as $\ket{0^n}$), the measurement results of single qubits are either deterministic or uniformly random. For $C\ket{0^n}$, and after choosing the measurement outcome for each of the $k$ qubits uniformly randomly, the output probability of an input $z\in \{0, 1\}^n$ is $|\langle z | C | 0^n\rangle|^2 = 2^{-k}$ \cite{KerznerClifford}.



\noindent{\bf Stabilizer and Extended Stabilizer} (IBM Quantum, Qiskit) \cite{IBMQuantum2021, QiskitDocumentation}: 
Stabilizer is a Clifford circuit simulator for noisy or ideal simulations up to 5000 qubits.
As an extension to Stabilizer, Extended Stabilizer has support for ideal modeling up to only 63 qubits.
With the addition of the $T$ gate, this simulator becomes universal for the aptly named \textit{Clifford+T} circuits.
Such circuits run polynomial in the number of Clifford gates and exponential in the number of $T$ gates \cite{bravyi2016clifford}.
The traditional, naive way to simulate Clifford+T circuits is to apply stabilizer decomposition by gadgetizing the $T$ gate into a small Clifford circuit containing a $CNOT$, an $S$, a magic state $\ket{A} = \frac{1}{\sqrt{2}}(\ket{0} + e^{i\pi / 4}\ket{1})$, and a measurement to classical control. The difficulty is then with the addition of $\ket{A}$ to the input state $\ket{0^n}\ket{A^{\otimes m}}$ where $n$ is the number of qubits and $m$ is the number of magic states. This new input state must be approximated by the simulator as the superposition of stabilizer states which are states generated by Clifford gates. This algorithm scales linearly in the number of stabilizer states with the goal of choosing the minimum number of necessary stabilizer states to accomplish the transformations.
Qiskit, however, uses a more recent, novel approach using sparsification and low-rank stabilizer decomposition \cite{bravyi2019stabilizerdecomposition} which also incorporates more than just $T$ gates \cite{gosset2019stabilizerdecomposition}.

\noindent{\bf cirq.CliffordSimulator} (Cirq, Google) \cite{cirq_developers_2022_7465577, CirqDocumentation}:
Google's Cirq features an efficient simulator tailored for Clifford circuits, leveraging the principles of the Gottesman-Knill theorem \cite{GottesmanPhD}. 
This simulator allows for a range of permitted operations, including X, Y, Z, H, S, CNOT, CZ gates, along with measurements in the computational basis.
The quantum state is described in two distinct forms: firstly, in terms of stabilizer generators, which consist of a set of $n$ Pauli operators $S_1, S_2, ..., S_n$ such that $S_i \ket{\psi} = \ket{\psi}$.
Additionally, this implementation is built on the framework established by Aaronson and Gottesman in 2004 \cite{aaronson2004improved}.
Secondly, the simulator supports the CH-form, as defined by Bravyi et al. in 2018 \cite{gosset2019stabilizerdecomposition}.
The CH-form is analogous to the stabilizer tableaux with the inclusion of the global phase.
This representation not only manages the overall phase but also provides access to the amplitudes of the state vector.

\noindent{\bf CHP} \cite{aaronson2004improved}: 
CHP is a simulator of quantum stabilizer circuits formed by CNOT, Hadamard, and $\pi/2$ Phase gates (CHP). CHP provides an environment for the simulation of stabilizer circuits with a large number of qubits (thousands) and frequent measurements with the goal of having a highly efficient simulator compared with general purpose simulators. CHP also enables the design and debugging of quantum error-correction architectures as well as numerical modeling of highly entangled systems. Designers can specify stabilizer circuits/algorithms in a simple assembly language. To enable bitwise measurement, CHP provides a single-qubit measurement gate. To experimentally evaluate CHP, its developers generate random stabilizer circuits with 200 to 3200 qubits. CHP is limited by quadratic growth of the number of bits needed to simulate a circuit with n input qubits.

\noindent{\bf Stim} (Google) \cite{gidney2021stim}:  Stim makes three improvements to the CHP simulator: vectorization, reference frame sampling, and stabilizer tableau inversion.
The vectorization improvement utilizes 256-bit wide AVX instructions.
Stim utilizes \textit{Pauli frames}: Pauli products where the global phase is ignored.
In other words, a Pauli frame stores information about each qubit, including whether the qubit's bit or phase has flipped relative to some reference state.
Stim propagates Pauli frames throughout the circuit which are applied to Clifford operations to generate circuit samples.
Multiple Pauli frames are operated on simultaneously in parallel SIMD operations.
With the reference frame sampling method, Stim first gets the initial reference sample using general stabilizer simulation.
This initial reference sample is a single sampling of the circuit that is used as a reference for generating batches of additional samples quickly.
Once the initial reference sample is acquired, no additional expensive stabilizer tableau simulation is needed.
All other samples can be gathered by propagating Pauli frames through the initial sample.
The author touts an impressive constant cost per gate rather than linear or quadratic.
In CHP, a tableau of binary variables is used to represent a state; this includes destabilizer generators, stabilizer generators, Pauli matricies, and phases.
The stabilizer tableau inversion used by Stim is a method in which the tableau is inverted such that measurements that commute with stabilizers take linear time rather than quadratic time.
The intuition is that with error correcting codes, for example, measurements are often redundant and redundant measurements would commute.
The simulator was run with several benchmarks: bulk sampling a surface code circuit, sampling a surface code circuit, sampling a Bacon-Shor code circuit, sampling a randomly generated circuit, and sampling multi-level S state distillation.
The surface code circuits and Bacon-Shor code circuits are error tolerant architectures for solid-state QCs \cite{fowler2012surfacecodes, gidney2019magicstatefactories}.
Within 15 seconds, Stim can analyze a distance 100 surface code circuit that contains 20K qubits, 8 million gates, along with 1 million measurements. 

\noindent{\bf Clifford Expansion} (HybridQ, NASA) \cite{HybridQ}:  HybridQ applies several optimizations to the density matrix simulation of stabilizer circuits. The first is that HybridQ compresses multiple gates together rather than applying smaller gates sequentially. This approach reduces density matrix "branches" by avoiding redundant branches. The second optimization also reduces the number of branches that need simulation by performing a depth-first expansion on Pauli strings (a tensor product representation of a multi-qubit Pauli operator such that each single qubit Pauli operator acts on a different qubit). Parallelization is available by distributing branches among threads through a breadth-first expansion. Message Passing Interface (MPI) is also utilized for distributed execution.

%% file: framework.tex
\section{Development Frameworks}
\label{sec:framework}

As quantum computing technology advances, several software frameworks have been developed to facilitate the development of quantum algorithms and circuits. These frameworks provide high-level abstractions that make it easy to write, manipulate, optimize, and run quantum circuits, as well as provide simulators to simulate the circuits on classical computers. In this section, we discuss several popular quantum software frameworks, including IBM Quantum Composer, Amazon Braket, TensorFlow Quantum, Cirq, Strawberry Fields, and HybridQ.
The constituent simulators that makeup these frameworks are categorized and discussed in previous sections.

\noindent{\bf IBM Quantum Composer} \cite{IBMQuantum2021}: 
The IBM Quantum Composer is a drag-and-drop graphical circuit designer with visualizations.
Some supported visualizations include the Bloch sphere, and visualizations for quantum probabilities and amplitudes.
IBM Quantum Composer includes the QASM general-purpose simulator that encompasses multiple simulation methods: statevector, density matrix, stabilizer and extended stabilizer, and matrix product state.
These methods are described throughout the paper.
By default, QASM will choose the best simulation method for the given circuit, but a specific method can always be chosen \cite{qiskitQASMdoc}.
In terms of development, IBM Quantum Composer has a scripting mode that is also available for programming in the Qiskit or OpenQASM language for simultaneous development using scripts and GUI. In addition to being open source, Qiskit is developed and maintained by IBM. As such, IBM Quantum retains all the benefits associated with Qiskit such as continuous updates. For example, a cache blocking technique \cite{doi2020cacheblocking} was added in 2021 that greatly improves the scalability of parallel simulation by inserting swap gates to decrease data movements. Through the companion IBM Quantum Lab, users can run their circuits on IBM's quantum computers. The IBM Quantum framework makes its various simulators available to run on IBM Cloud (limited to 8192 shots) or locally.

\noindent{\bf Amazon Braket} \cite{AmazonBraket, AmazonBraketDevices}: 
Amazon Braket supports a fully managed Jupyter notebook development environment using the Amazon Braket Python SDK.
Amazon's several quantum computers with various processors are available to run circuits through the SDK.
Amazon Braket supports the following simulators available to run as part of Amazon's managed simulation or locally: state vector, density matrix, analog Hamiltonian, and tensor network.

\noindent{\bf TensorFlow Quantum} (Google) \cite{broughtonTensorFlowQuantum2020, TensorFlowQuantumDocumentation}: 
TensorFlow Quantum (TFQ) is an open source quantum variation on TensorFlow; both developed by Google. This framework supports high-performance circuit simulators including their native qsim and qsimh as well as Cirq simulators. Tools are provided to support interleaving TFQ with logic developed within the Cirq framework. TFQ merges quantum algorithms with machine learning such that TFQ may use back-propagation compatible gradient calculations to generate a parameterized quantum circuit. Circuits generated this way are optimized for running on NISQ-era hybrid quantum computers.

\noindent{\bf Cirq} (Google) \cite{cirq_developers_2022_7465577, CirqDocumentation}:
Cirq is an open-source Python library developed by the Quantum AI team at Google to facilitate the development of quantum algorithms and circuits on NISQ machines.
It provides a high-level abstraction layer that makes it easy to write, manipulate, optimize, and run quantum circuits all within its framework.
In addition to supporting gate-level operations, Cirq supports higher-level operations such as parallel and sequential operations, subroutines, and measurements. Cirq's API is designed to be simple and intuitive, making it easy for developers to get started with quantum computing without needing an extensive background in physics or mathematics.
Cirq includes several native simulators, such as cirq.Simulator and cirq.DensityMatrixSimulator, that support both pure and noisy quantum states. Additionally, Cirq supports several externally developed high-performance simulators, such as qsim, qsimh, quimb, and qFlex, that can be used to simulate large circuits that may be difficult to simulate using the native simulators provided by Cirq.
Cirq's native simulators are continually updated with performance improvements. For example, the performance of circuits with disconnected subsystems was drastically improved by dynamically factorizing the simulated state into non-entangled subsets of qubits \cite{CirqReleases}.
Cirq is compatible with Google's Quantum Computing Service that gives customers access to Google's quantum computing hardware. To do so, the circuit must be properly prepared and optimized for improved performance.
Several techniques are used: conversion to target gateset, ejection of single-qubit operations, alignment of gates in moments, and an error mitigation technique called spin echos \cite{CirqDocumentationSpinEchoes}.


\noindent{\bf Strawberry Fields} \cite{killoranStrawberryFields2019, BromleyStrawberryFieldsApplications2020, StrawberryFieldsDocumentation, StrawberryFieldsGitHub}: Straberry Fields is an open-source quantum programming architecture for photonic quantum computers with support for simulation, optimization, and quantum machine learning of continuous-variable circuits. Strawberry Fields supports an API with a custom purpose-built language called Blackbird, several backends built in NumPy and TensorFlow, and three built-in simulators.

\noindent{\bf HybridQ} (NASA) \cite{HybridQ}:  HybridQ is an open-source, high performance framework built for distributed computing developed by NASA in Python and C++. Parallelization and scalability are at the core of this framework's three simulator types: state vector, stabilizer, and tensor network contraction. The simulators support the Message Passing Interface (MPI) for simulating on distributed systems, and noisy and noiseless simulation. For ease of development, HybridQ is integrated with SymPy, a symbolic mathematics Python library, such that symbols can be used to parameterize gates where the symbols can be defined later. To maximize code recycling, HybridQ gives gates properties which are independently developed. Gates are defined by these properties such that any improvement to a property is propagated to all other gates with that property.

\noindent{\bf Azure Quantum Development Kit (QDK)} (Microsoft) \cite{Azurequantum}: The Azure Quantum Development Kit created by Microsoft is an open source application which operates on and has been integrated with Azure Quantum offering a surplus of tools for quantum development. Algorithms are developed in Q\# a unique scripting language to the Microsoft environment that provides access to Quantum computers on top of state of the art quantum simulators. Q\# was developed in order to improve the scalability of other simulators such as Liqui$\mid>$. Q\# describes quantum algorithms in terms of expressions instead of focusing on a gate by gate description, and this allows for a classical approach to writing quantum algorithms. Microsoft Quantum Development Kit provides access to five unique quantum simulators as well as original quantum hardware. In order these simulators are: full state simulator, sparse simulator, trace-based resource estimator, Toffoli simulator, and noise simulator. The simulators of note include the sparse simulator and Toffoli simualtor. This framework strives to provides some of the best simulators available while also focusing heavily on the ability to teach and research quantum computation.

%% file: challenges.tex
\section{Challenges and Roadmap}
\label{sec:rdmp}


This section discusses some of the most important challenges before the development of quantum simulators and their wider use in research and education. First and foremost, we note that despite significant progresses made in the past decade, the {\it scalability} of simulating large quantum circuits (in terms of their width and depth) on classic machines remains the main challenge. Figures 3-5 summarize different techniques for optimizing the performance of quantum simulators when dealing with exponential size matrices and vectors of complex values. (Such matrices and vectors respectively capture quantum transformations and states.) These methods can be classified into matrix/vector partitioning, parallelization, instruction set vectorization, SIMD-based approaches, and gate fusion. Since no single approach is a silver bullet to the scalability problem, existing approaches often combine a few optimization techniques (see Figures 3-5). Moreover, some approaches utilize existing hardware capabilities for SIMD and vectorized computation of quantum transformations. 
In addition to aforementioned approaches, we believe that there are other ways for improving the scalability of quantum simulators. 
\begin{itemize}

\item
First, the more efficient execution of quantum computations on classical machines warrants \textbf{custom-designed processors}.
Such processor designs can use available hardware technologies.
Inspired by the challenges of representing quantum states and transformations in a classic computational model, researchers may focus on developing processors/accelerators that can perform quantum transformations highly efficiently.
For the development of custom processors, Application-Specific Integrated Circuits (ASIC) can be cost prohibitive to researchers; however, Field-Programmable Gate Arrays (FPGA) are an accessible alternative.
Pilch and Długopolski \cite{pilch2019FPGA} designed an FPGA modeled around the requirements of general quantum circuit simulation with hardware-analogy to the physical reality of quantum computation.
For the most flexibility, High-Performance Re-configurable Computers (HPRC) combines FPGAs with CPUs.
El-Araby et al. \cite{el-araby2023HPRC} configured an HPRC machine to perform a quantum Haar transform (a wavelet transform similar to the Fourier transform) \cite{fijany1998quantum, chui1992wavelets} on an image. They found significant speedup and space reduction compared to Qiskit.
Such works show that FPGA integration with existing simulation frameworks would yield significant speedup. However, there are serious challenges that should be addressed such as memory-I/O bottleneck in such systems and the emulation of quantum measurement.



\item Another area for future work is to devise a {\bf repository of circuit-specific simulators} where a collection of simulators are developed, each appropriate for the simulation of specific classes of quantum circuits/algorithms. For example, one can have simulators that are highly efficient for stabilizer circuits (e.g., CHP \cite{aaronson2004improved}), low-depth but high-width circuits, and high-depth but low-width circuits. Depending on the quantum circuit at hand, one can retrieve a simulator that is specific to that circuit, thereby ensuring high efficiency and scalability for the circuits in that family. This approach is contrary to developing generic simulators that can simulate any kind of quantum circuits. 

\item A third avenue for future work includes the development of {\bf hardware-specific simulators}. For example, while there are many simulators that can utilize the processing power of GPUs, they are not necessarily {\em designed} for the GPU architecture; rather they use existing APIs/libraries developed for performing matrix-based calculations on GPUs. As such, we believe that hardware-specific simulators can greatly enhance the scalability of simulated quantum computations on classic machines. 

\item The development of {\bf domain-specific simulators} includes another area that needs further attention. For example, there are a few quantum simulators and libraries for the modeling of molecular structures \cite{stair2022qforte,fan2022q}. It is desirable that quantum computer scientists closely collaborate with domain experts of other fields of science towards developing highly efficient quantum simulators customized for the challenges of the domain of interest (e.g., Chemistry, Biology, etc.). This is an area that is open to more research with the objective of developing highly efficient domain-specific quantum simulators.

\item Finally, {\bf custom-designed data structures} are at the core of tackling the quantum state/transformation representation. Developing space and time-efficient data structures that can provide a small memory footprint for quantum state and can enable time-efficient quantum transformations is an important area of further investigations. For instance, some researchers \cite{burgholzer2022simulation,wille2022tools,burgholzer2023design} have achieved significant improvements through customizing Quantum Decision Diagrams (QDDs). Nonetheless, finding a variable ordering that provides a compact and conical representation in QDDs is by itself a hard problem.

\end{itemize}

%% file: concl.tex
\section{Conclusions}
\label{sec:concl}


This paper presented a literature study and classification of methods of simulating quantum algorithms/circuits on classic machines. The major findings of this study includes the following. First, we realized that most efficient and actively maintained simulators have been developed after 2010. Second, we classified the most important simulation methods in four categories: Schrodinger-based, Feynman path integrals, Heisenberg-based, and hybrid methods. While the most commonly used simulators are Schrodinger-based, there are a few efficient simulators in other categories. Quantum frameworks also provide more flexibility in terms of the choice of simulators/simulation method. We also classified and found out that some state-of-the-art simulators utilize a combination of software and hardware techniques to scale up the simulations. Further, we identified some directions for future research for advancement of quantum simulators in education and research.